\documentclass[twocolumns]{aa}
\usepackage{graphicx}
\def\ls{{_<\atop^{\sim}}}
\def\gs{{_>\atop^{\sim}}}
\def\cgs{ ${\rm erg/cm}^{2}/{\rm s}$ } 

\begin{document}
%
\title{Faint high-redshift AGN in the Chandra Deep Field South: the
  evolution of the AGN luminosity function and black hole
  demography}

\author{F. Fiore\inst{1}
\and
S. Puccetti\inst{2}
\and
A. Grazian\inst{1}
\and
N. Menci\inst{1}
\and
F. Shankar\inst{3}
\and
P. Santini\inst{1}
\and
E. Piconcelli\inst{1}
\and
A.M. Koekemoer\inst{4}
\and
A. Fontana\inst{1}
\and
K. Boutsia\inst{1}
\and
M. Castellano\inst{1}
\and
A. Lamastra  \inst{1}
\and
C. Malacaria\inst{1,5}
\and
C. Feruglio \inst{6}
\and
S. Mathur \inst{7}
\and
N. Miller \inst{8}
\and
M. Pannella \inst{9}
}

\institute{
Osservatorio Astronomico di Roma, via Frascati 33, I00040 Monteporzio Catone, Italy
              \email{fabrizio.fiore@oa-roma.inaf.it}
\and
ASI Science Data Center, Italy
\and
Max-Planck-Institut fur Astrophysik, Germany
\and
Space Telescope Science Institute, USA
\and
Universita' di Roma La Sapienza, Italy
\and
Institut de RadioAstronomie Millimétrique, France
\and
Ohio State University, USA
\and
University of Maryland, USA
\and
CEA Saclay, Service d'Astrophysique, France
}

\date{August 25, 2011}

\abstract  
{We present detection and analysis of faint X-ray sources in
the Chandra deep field south (CDFS) using the 4 Msec Chandra
observation.}
{We put new constraints on the z=3--7 AGN luminosity function, its cosmological evolution and
high-redshift black hole and AGN demography.}
{We use a new detection algorithm, using the entire
3-dimensional data-cube (position and energy), and use a targeted
search at the position of high-z galaxies in the GOODS-South survey.}
{This optimized technique results in the identification of 54 z$>$3
  AGNs, 29 of which are new detections. Applying stringent
  completeness criteria, we derive AGN luminosity functions in the
  redshift bins 3--4, 4--5 and $>$5.8 and for 42.75$<$log L(2--10
  keV)$<$44.5. We join this data with the luminous AGN luminosity
  functions from optical surveys and find that the evolution of the
  high-z, wide luminosity range luminosity function can be best
  modeled by pure luminosity evolution with $L^*$ decreasing from
  $6.6\times 10^{44}$ ergs/s at z=3 to $L^*=2\times 10^{44}$ ergs/s at
  z=6. We compare the high-z luminosity function with the prediction
  of theoretical models using galaxy interactions as AGN triggering
  mechanism.  We find that these models are broadly able to reproduce
  the high-z AGN luminosity functions. A better agreement is found
  assuming a minimum Dark Matter halo mass for black hole formation
  and growth.  We compare our AGN luminosity functions with galaxy
  mass functions to derive high-z AGN duty cycle using observed
  Eddington ratio distributions to derive black hole masses. We find
  that the duty cycle increases with galaxy stellar mass and redshift
  by a factor 10-30 from z=0.25 to z=4-5.  We also report on the
  detection of a large fraction of highly obscured, Compton thick AGN
  at z$>3$ ($18^{+17}_{-10}\%$). Their optical counterparts do not
  show any reddening and we thus conclude that the size of the X-ray
  absorber is likely smaller than the dust sublimation radius. We
  finally report the discovery of a highly star-forming galaxy at
  z=3.47, arguing that its X-ray luminosity is likely dominated by
  stellar sources. If confirmed, this would be one of the farthest
  objects in which stellar sources are detected in X-rays.}
{}

\authorrunning {Fiore et et al.}
\titlerunning {High-z AGN in the CDFS}

\maketitle

%

\section{Introduction}

The study of high redshift AGN holds the key to understand early
structure formation and probe the Universe during its infancy.  High-z
AGN have been extensively used to investigate key issues such as the
evolution of the correlations between the black hole mass and the
galaxy properties (see e.g \cite{lamastra:2010} and references
therein), the AGN contribution to the re-ionization, heating of the
Inter-Galactic Matter and its effect on structure formation
(e.g. \cite{giallongo:1997,malkan:2003,hopkins:2007,cowie:2010,
  boutsia:2011}).  However, there are other fundamental issues that
can be tackled by studying high-z AGN: a) scenarios for the formation
of the black hole (BH) seeds, which will eventually grow up to form
the super-massive black holes (SMBHs) seen in most galaxy bulges. Two
main scenarios have been currently proposed, the monolithic collapse
of $10^4-10^5$ M$\odot$ gas clouds to BH
(\cite{lodato:2006,lodato:2007,begelman:2010,volonteri:2010b}), or
early generation of $\sim100$ M$\odot$ BH produced by the supernova
explosions of the first Pop III stars (\cite{madau:2001}).  b)
Investigate the physics of accretion at high-z: is BH growth mainly
due to relatively few accretion episodes, as predicted in hierarchical
scenarios (see e.g. \cite{dotti:2010} and references therein), or by
the so called chaotic accretion (hundreds to thousands of small
accretion episodes, \cite{king:2008})? The two scenarios predict
different BH spin distributions, and thus different distributions for
the radiative efficiency.  c) Being BH the structures with the fastest
(exponential) growth rate, they can be used to constrain both the
expansion rate of the Universe and the growth rate of the primordial
perturbation at high-z, i.e. competing cosmological scenarios
(\cite{fiore:2010,lamastra:2011}).  d) The slope of the high-z AGN
luminosity function and of the SMBH mass function strongly depend on
the AGN duty cycle, and therefore their measure can constrain this
critical parameter. In turn, the AGN duty cycle holds information on
the AGN triggering mechanisms.  The evaluation of the evolution of the
AGN duty cycle can thus help in disentangling among competing
scenarios for AGN triggering and feeding. In this paper we discuss in
more detail this last issue.  Two main scenarios for AGN triggering
and feeding have been investigated so far: galaxy encounters
(\cite{barnes:1996,cavaliere:2000,menci:2006,menci:2008}), and
recycled gas produced by normal stellar evolution in the inner bulge
(\cite{ciotti:2007,ciotti:2010,cen:2011}). These make different
predictions about the AGN time scale and duty cycle. In the former
models nuclear activation follows galaxy interactions. Since these are
naturally more frequent in the past, when, at the same time, more gas
is available for nuclear accretion since it has not yet been locked in
stars, this scenario predicts a strong increase of the AGN duty cycle
with the redshift. In the recycled star gas scenario the AGN timescale
is much longer, up to Gyr, although with decreasing Eddington ratio. A
less extreme variation of the AGN duty cycle with the redshift is thus
expected. One of the goals of this paper is to assess what can be
already done with present AGN surveys to distinguish between these
competing scenarios, and which kind of surveys should be planned to
best assess the main AGN triggering and feeding mechanism.

Large area optical surveys such as the SDSS, the CFHQS, and the NOAO
DWFS/DLS have already been able to discover large samples of z$>4.5$
QSOs (e.g. \cite{richards:2006,glikman:2011}) and about 50 QSOs at
$z>5.8$ (e.g. \cite{jiang:2009,willott:2010a}). The majority of these
high-z AGN are broad line, unobscured, high luminosity AGN. They are
likely the tips of the iceberg of the high-z AGN population.  Lower
luminosity and/or moderately obscured AGN can, in principle, be
detected directly in current and future X-ray surveys.  Dedicated
searches for high-z AGN using both deep and wide area X-ray surveys
and a multi-band selection of suitable candidates can increase the
number of high-z AGN by a factor $>10$. In particular, it should be
possible to find hundreds rare high-z, high luminosity QSOs, in both
the all sky and deep eROSITA surveys (the 0.5-2 keV flux limit of the
all sky survey being the order of $10^{-14}$ \cgs, while that of the
deep survey, covering hundreds deg$^2$, should be 2-3 times deeper)
with a selection function much less biased than optical surveys
(recall that $\tau_X/\tau_{opt}\approx (1+z)^{-3.5}$, where $\tau_X$
and $\tau_{opt}$ are the optical depths in the observed-frame X-ray
and optical bands, assuming no evolution of the dust-to-gas ratio). To
constrain the faint end of the high-z AGN luminosity function, and
therefore the shape of the luminosity function and of the SMBH mass
function, we need to best exploit current and future deep
surveys. Unfortunately, so far the number of X-ray selected AGN at
z$>3$ in deep Chandra and XMM surveys is only of a few tens, and
$\sim$half a dozen at z$>4.5$ (see
e.g. \cite{brandt:2004,fontanot:2007,brusa:2009a,brusa:2009b,civano:2011}).
The difficulty in detecting directly high-z AGN in X-ray surveys
suggests an alternative approach. Looking at the X-ray emission at the
position of known sources allows one to use a less conservative
threshold for source detection than in a blind search, and therefore
to reach a lower flux limit. Furthermore, a highly optimized X-ray
analysis tool is deemed mandatory to fully exploit the richness of the
data produced by Chandra and XMM.  To this purpose we developed a new
X-ray detection and photometry tool (dubbed {\em ephot}, see the
Appendix).  In this paper we search for X-ray emission at the position
of candidate high-z galaxies selected in the HST/WFC3 ERS (Early
release science) area and in the GOODS area of the Chandra Deep Field
South (CDFS) by using {\em ephot}. We use the new 4 Msec Chandra
dataset, which represent the most sensitive X-ray exposure ever
achieved. We first study selection effects affecting our candidate
high-z AGN, and then join our CDFS samples to other X-ray and
optically selected AGN samples at the same redshift to build AGN
luminosity functions at z$>3$ over several luminosity decades. By
assuming appropriate bolometric corrections and appropriate
distributions of $\lambda=L_{bol}/L_{Edd}$ we convert these luminosity
functions in SMBH mass functions. We then evaluate the stellar mass
functions of {\it active} galaxies by converting the SMBH mass to
galaxy stellar mass by using the {\em Magorrian} relationship
(\cite{ferrarese:2005, shankar:2009a} and references therein).
Finally we estimate the AGN duty cycle as a function of the redshift
by dividing the stellar mass function of {\it active} galaxies by the
stellar mass function of {\it all} galaxies at the same redshift.  The
AGN duty cycle as a function of the redshift is then compared with the
expectation of models for structure formation using galaxy interaction
as AGN triggering mechanism.  A $H_0=70$ km s$^{-1}$ Mpc$^{-1}$,
$\Omega_M$=0.3, $\Omega_{\Lambda}=0.7$ cosmology is adopted
throughout.

\section{Candidate high-z galaxy samples}

We built a catalog of z$>3$ galaxy candidates in the GOODS south and ERS fields including:

\begin{itemize}
\item
all galaxies with high quality spectroscopic redshift $>3$;
\item
all galaxies with 68\% limit of photometric redshift $>3$;
\item
all galaxies with B-V vs V-I or V-H colors suggesting 3.5$<$z$<$4.4;
\item
all galaxies with V-I vs I-Z or I-H colors suggesting 4.4$<$z$<$6.0;
\item
all galaxies with I-Z vs Z-Y or Z-Y vs Y-J colors suggesting z$>$6;
\item
all galaxies with Y-J vs J-H colors suggesting z$>$8.
\end{itemize}

\subsection{GOODS field}

We use the GOODS-MUSIC catalog (\cite{grazian:2006,santini:2009}),
based on the mosaics reduced as described in Koekemoer et al. (2011),
see also Windhorst et al. (2011), which includes galaxies selected in
the HST/ACS $z$ band and Spitzer/IRAC 4.5$\mu m$ band.  190
GOODS-MUSIC galaxies have a high quality optical or near-infrared
spectrum confirming that they are z$>3$ galaxies, 26 have a
spectroscopic confirmation at z$>5$.  4284 galaxies without high
quality spectroscopic redshift z$_{spec}$ have photometric redshift
z$_{phot}>3$, considering the 68\% upper limit.  To this sample, we
added 171 galaxies without high quality z$_{spec}$ and z$_{phot}<3$
that lie in the 3.5$<$z$<$4.4 region of the ACS B-V vs V-I diagram and
143 galaxies that lie in the 4.4$<$z$<$6.0 region of the ACS V-I vs
I-Z diagram.  The total number of high-z galaxy candidates considered
is thus 4788.  About 10\% of these galaxies lie within a few arcsec
from an X-ray source previously identified with a low-z AGN or galaxy
or galaxy cluster. These objects were conservatively excluded from the
list, leaving us with about 4300 high-z galaxy candidates.

\subsection{ERS field}

We use the GOODS-ERS catalog (\cite{grazian:2010,santini:2011}), which
includes 2291 galaxies selected in the HST/WFC3 H band and with
photometric redshift $>$3. 42 of these objects have a confirmed
spectroscopic redshift (only 2 with z$>5$). Again, about 10\% of these
galaxies lie close to bright X-ray sources, thus limiting the ERS
candidate high-z galaxy sample to about 2000 galaxies.

There is of course an overlap between the GOODS-ERS and the GOODS-MUSIC
catalogs (the GOODS field covers nearly completely the ERS field). Of
the 2291 GOODS-ERS, z$>3$ galaxies 570 are present in the GOODS-MUSIC
catalog and in 420 cases the GOODS-MUSIC photometric redshift is $>3$
Therefore, these objects are present also in the GOODS-MUSIC 
high-z galaxy candidate sample.

\subsection{Additional samples}

We also considered 330 z$>6$ galaxy candidates from the lists of
\cite{bouwens:2006,castellano:2010,mclure:2010,mclure:2011,bouwens:2010}, which are
not included in the above two lists.
  
\section{Multiwavelength properties of High-z AGN and galaxies in the CDFS.}

\begin{figure}
\centering
\includegraphics[width=8.5cm]{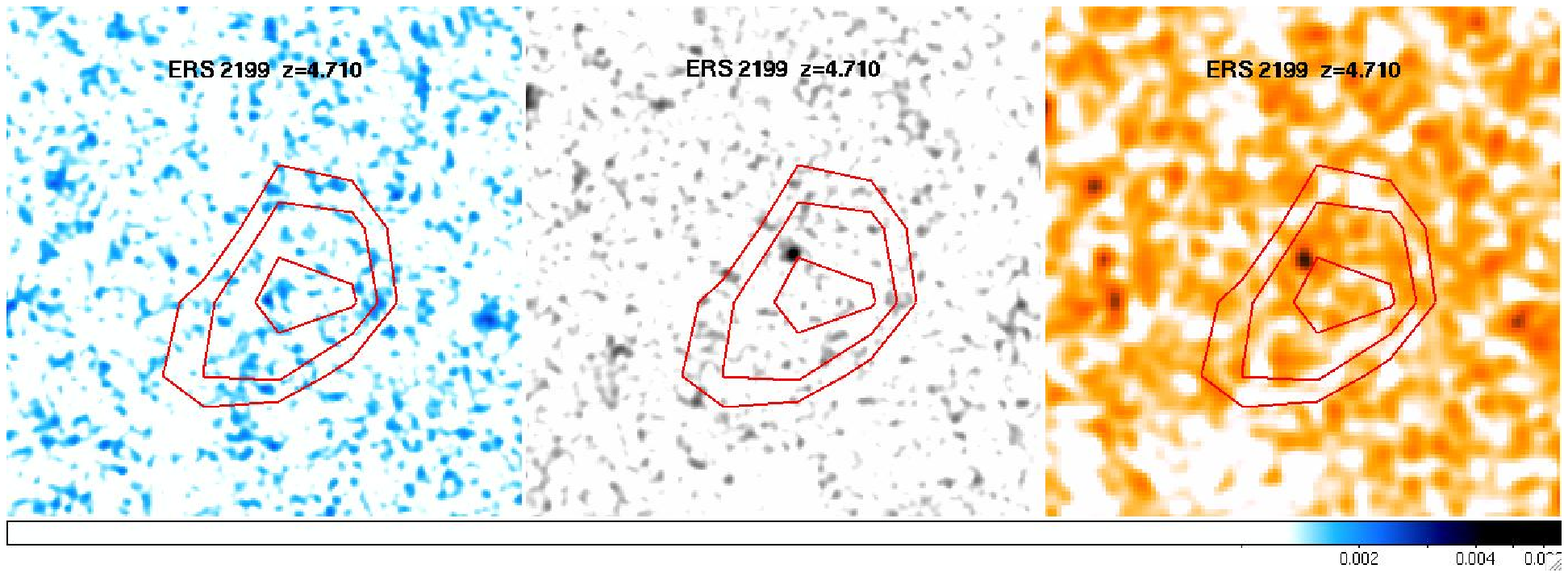}
\includegraphics[width=8.5cm]{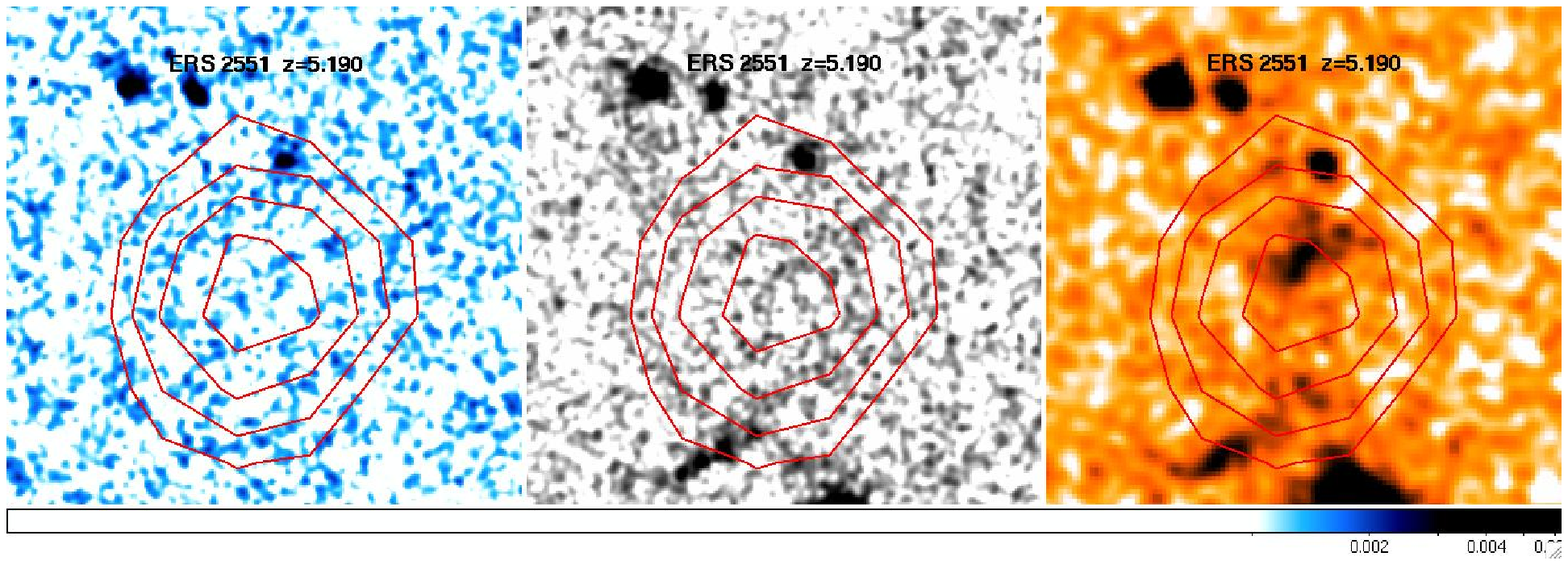}
\includegraphics[width=8.5cm]{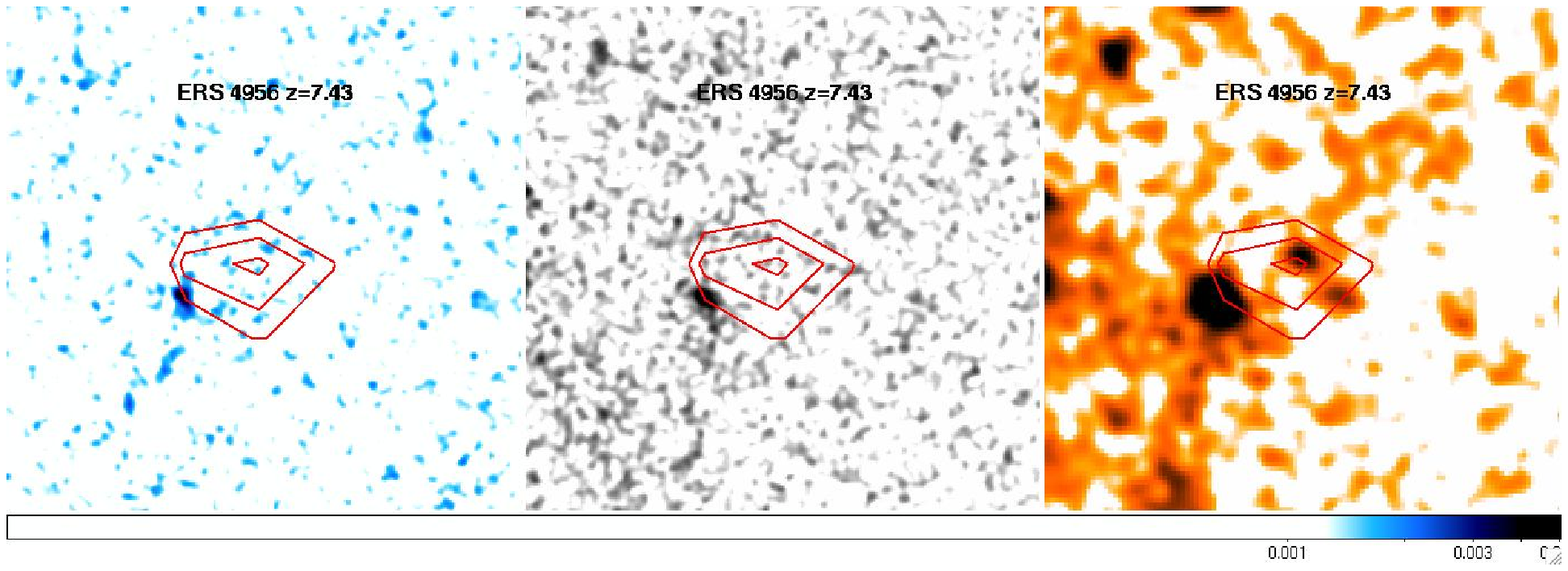}
\caption{HST ACS B band (left panel), $z$ band (central panel) and HST
  WFC3 H band (right panel) images of three Chandra-GOODS-ERS sources
  with overlaid X-ray contours in the 0.7-2.7 keV band (E2199),
  0.7-1.9 keV band (E2551) and 0.4-1.5 keV band (E4956)}
\label{ima}%
\end{figure}

To extract statistically quantitative information from a given sample
of ``detected'' sources in an X-ray image, we need to assess its
reliability (i.e. number of spurious detections) and completeness.  To
both purposes we run a series of extensive detection runs on simulated
data, see the Appendix for details.  We choose a detection threshold
corresponding to 1 spurious detection in 5000 candidates, i.e. about 1
spurious detection in the GOODS-MUSIC sample and $<$1 spurious
detection in the GOODS-ERS sample. We detect 17 sources at z$>3$ from
the GOODS-ERS catalog and 41 from the GOODS-MUSIC catalog within this
threshold.  As an example, Fig. \ref{ima} show the B, $z$ and H band
images of three Chandra-GOODS-ERS sources.  Table 1 and 2 give
position, redshift, X-ray band that optimizes the detection, 0.5-2 keV
flux, z band and H band magnitudes and 4.5$\mu$m fluxes for the
GOODS-ERS and GOODS-MUSIC X-ray detected sources, along with a few
sources just below detection threshold (4 from the ERS-GOODS sample, 2
from the GOODS-MUSIC).  We do not detect any of the candidate z$\gs6$
galaxies in
\cite{bouwens:2006,castellano:2010,mclure:2010,mclure:2011,bouwens:2010}.
One of these galaxies has an X-ray counterpart just below threshold
(also reported in Table 2). At the position of this source there are
11 counts in the band 0.4-0.8 keV ($\sim3$ of which are background counts)
in a 2 arcsec radius region. If the 0.4-0.8 keV emission is due to
iron K$\alpha$ emission at 6.4 keV the redshift would be $\gs8$, which
is inconsistent with the detection in the $z$ band. Because of these
uncertainties we do not discuss further this source.

Five Chandra-GOODS-ERS sources are present in the 2 Msec direct
detection list of \cite{luo:2008} (Luo08 hereafter, search within a
circle of 1 arcsec radius). Three of them have z$_{phot}$ in
\cite{luo:2010} (Luo10 hereafter) consistent with those in the
GOODS-ERS catalog, while in one case the Luo10 z$_{phot}$ is just
below 3. One Chandra-GOODS-ERS source is confused/blended in the Luo08
catalog.  Eight Chandra-GOODS-ERS sources are present in the 4 Msec
direct detection list of \cite{xue:2011} (Xue11 hereafter). Two of the
three Xue11 sources not in the Luo08 catalog have a photometric
redshift, and this is $>3$ in both cases.  16 Chandra-GOODS-MUSIC
sources are present in the Luo08 list, 13 with photometric or
spectroscopic redshift consistent with those in the GOODS-MUSIC
catalog.  Three additional sources are present in the Xue11 catalog,
two of them have a photometric redshift, and this is consistent with
that in the GOODS-MUSIC catalog. Source M208 was not detected by
Luo08, while it is now detected by Xue11. It is extensively discussed
in \cite{gilli:2011}.

The accuracy of the photometric redshifts is greatly improved thanks
to the near infrared bands covered by WFC3 (see Table 1 and 2). 
The WFC3 coverage also increases the high-z candidate galaxy
X-ray detection rate. The ERS area is less than one third of the full
GOODS area, and therefore, based on the Chandra-GOODS-ERS detections,
one would expect $\gs54$ detections in GOODS-MUSIC, while only 41 have
been actually found. Only half of the Chandra-GOODS-ERS detections
have a counterpart in the GOODS-MUSIC catalog.  31\% and 40\% of the
Chandra-GOODS-MUSIC detections are also present in the Luo08 2 Msec
and Xue11 catalogs.  28\% and 47\% of the Chandra-GOODS-ERS
detections are present in the same catalogs. We study in more detail
the different Chandra-GOODS-MUSIC and Chandra-GOODS-ERS selection
effects in section 3.3.

\begin{table*}
\caption{\bf Chandra detected GOODS-ERS z$>3$ galaxies}
\begin{tabular}{lccccccccl}
\hline
ID    &  RA        &  Dec        &  z    & $E_{min-max}$ & SNR & F(0.5-2keV)& zm & H & other ID$^a$ \\
      &  J2000     & J2000       &       & keV          &     &  \cgs      &    &   &     \\
\hline
E537   &  53.082535 &  -27.755245 & 4.29 (2.9-4.5) & 1.3-6.3 & 13.6 & 1.2E-16 & 25.48 &24.21 & L184, X275, M70139$^b$ \\
E737   &  53.050343 &  -27.752138 & 4.14 (4.0-4.4) & 1.0-2.0 &  3.4 & 4.5E-17 & 25.65 &24.94 & X177, M70131$^c$\\
E1312  &  53.016602 &  -27.744848 & 3.48 (3.4-3.6) & 0.9-2.1 &  5.6 & 1.0E-16 & 25.30 &24.39 & L68, X100, M13365$^d$, MY50634 \\
E1516  &  53.119888 &  -27.743042 & 4.29 (4.1-6.9) & 0.5-1.9 &  3.4 & 3.2E-17 & 27.99 &25.57 & X392, M70099\\
E1577  &  53.020576 &  -27.742151 & 3.462 spec.    & 1.9-6.3 &  3.7 & 9.5E-17 & 23.62 &23.35 & M13549,MY51260 \\
E1617  &  53.010651 &  -27.741613 & 3.38 (3.3-3.4) & 0.9-3.8 &  7.7 & 1.9E-16 & 26.33 &24.58 & L61$^e$,X90 \\
E2199  &  53.054756 &  -27.736839 & 4.71 (4.5-5.2) & 0.7-2.7 &  2.9 & 3.5E-17 & 27.15 &27.08 & \\
E2309  &  53.117855 &  -27.734295 & 3.61 (3.3-3.8) & 0.3-4.0 &  5.3 & 4.7E-17 & 23.81 &22.93 & L252, X386, M14078\\
E2498  &  53.006237 &  -27.734076 & 6.62 (5.9-7.3) & 1.0-5.3 &  5.0 & 5.1E-17 &-28.89 &26.61 & L57$^f$, X85 \\
E2551  &  53.158340 &  -27.733500 & 5.19 (5.1-5.5) & 0.7-1.9 &  4.3 & 5.0E-17 &-27.59 &25.00 & G557,L323,X521,M70091\\
E2658  &  53.055538 &  -27.732950 & 3.74 (3.5-4.1) & 0.4-6.0 &  3.8 & 4.2E-17 & 26.84 &26.85 & \\
E4956  &  53.037708 &  -27.711536 & 7.43 ($>$7.0)$^*$& 0.4-1.5& 3.2 & 6.3E-17 &-29.20 &26.97 & \\
E5165  &  53.038258 &  -27.709778 & 3.87 (3.6-4.2) & 0.5-2.0 &  3.3 & 7.0E-17 & 26.80 &26.94 & L100$^g$ \\
E6257  &  53.168419 &  -27.719418 & 3.22 (3.2-3.5) & 0.8-1.9 &  3.1 & 4.1E-17 & 24.64 &24.56 & X554, M15144/M15156$^h$, MY55184 \\
E7911  &  53.194885 &  -27.699949 & 4.86 (4.7-5.4) & 0.8-1.8 &  3.2 & 1.3E-16 & 99.00 &25.68 & \\
E8479  &  53.167572 &  -27.694883 & 3.67 (3.3-3.9) & 1.5-3.2 &  3.2 & 7.0E-17 & 26.56 &26.27 & MY58818 \\
E10247 &  53.131218 &  -27.685293 & 3.95 (3.6-4.0) & 0.5-4.5 &  3.1 & 3.2E-17 & 26.55 &26.59 & \\
\hline
E1611$^h$  &  53.079372 &  -27.741625 & 6.91 (6.0-7.4) & 0.8-1.7 &  2.6 & 2.5E-17 & 26.15 &23.75 & M70107\\
E7840$^h$  &  53.173592 &  -27.661259 & 3.68 (3.3-4.0) & 0.3-6.3 &  2.9 & 4.9E-16 & 25.04 &25.44 & MY64565\\
E8439$^h$  &  53.183681 &  -27.694748 & 4.91 (4.8-5.2) & 0.8-2.5 &  2.9 & 3.2E-17 & 99.00 &25.26 & \\
E9850$^h$  &  53.102806 &  -27.679691 & 3.93 (3.8-4.1) & 0.9-4.7 &  2.9 & 2.1E-16 & 26.36 &26.46 & \\
\hline
\end{tabular}

$^a$L=Luo08, M=GOODS-MUSIC, MY=GOODS-MUSYC Cardamone et al. 2010;X=Xue11;
$^b$GOODS-MUSIC z=2.28; $^c$GOODS-MUSIC z=2.33; $^d$no redshift in GOODS-MUSIC catalog; 
$^e$Luo10 z=2.123; $^f$Luo10 z=2.808;
$^g$Luo08 confused with optically bright galaxy; $^h$M15144 z=0.15 M15156 z=3.07;
$^*$secondary solution at z$\sim$2;$^h$ just below detection threshold.
\end{table*}

\begin{table*}
\caption{\bf Chandra detected GOODS-MUSIC z$>3$ galaxies}
{\footnotesize
\begin{tabular}{lcccccccccl}
\hline
ID    &  RA        &  Dec        &  z    & E$_{min-max}$ & SNR & F(0.5-2keV)& zm & H & F(4.5$\mu m$) & other ID$^a$ \\
      &  J2000     & J2000       &       & keV          &     &  \cgs      &    &   &  $\mu$Jy & \\
\hline
M208$^b$& 53.122036&  -27.938740 & 4.762 spec.     & 1.1-4.0 &  3.6 & 1.0E-16 & 24.99 & 23.42 & 3.0 & G618,X403,MY19471 \\
M387   & 53.088505 &  -27.933449 & 3.130 (2.4-3.5) & 0.5-1.3 &  2.7 & 6.7E-17 & 25.02 & 99.00 & 1.1 &            \\
M603   & 53.109451 &  -27.928093 & 3.360 (2.2-3.9) & 0.8-1.9 &  2.9 & 3.7E-17 & 25.82 & 25.09 & 0.5 & MY21391    \\
M2690  & 53.163532 &  -27.890448 & 3.620 (1.5-4.6) & 0.7-6.0 &  3.0 & 1.1E-16 & 25.67 & 24.96 & 9.1 & MY27345    \\
M3268  & 53.083511 &  -27.881855 & 3.300 (2.6-3.8) & 1.7-6.3 &  3.2 & 1.0E-16 & 26.46 & 26.47 & 0.2 & MY28191    \\
M3320  & 53.209370 &  -27.881086 & 3.470 spec      & 0.7-5.0 & 41.1 & 2.5E-15 & 25.29 & 99.00 & 4.5 & G159, L402,X645 \\
M3323  & 53.184639 &  -27.880917 & 3.471 spec.     & 0.5-2.1 & 12.7 & 4.5E-16 & 23.98 & 23.84 & 4.0 & G21,L369,X588   \\
M4302  & 53.174389 &  -27.867350 & 3.610 spec.     & 0.4-5.0 & 30.6 & 1.5E-15 & 22.43 & 22.44 & 9.6 & MY30943,G24,L350,X563 \\
M4417  & 53.097237 &  -27.865793 & 3.470 spec      & 0.5-3.2 &  3.1 & 2.2E-17 & 23.33 & 23.15 & 2.5 & MY31136    \\
M4835  & 53.078468 &  -27.859852 & 3.660 spec.     & 0.8-6.3 & 12.4 & 1.2E-16 & 25.08 & 99.00 & 7.3 & G263b,L180,X262 \\
M5390$^c$ &53.124371& -27.851633 & 3.700 spec.     & 0.7-6.3 & 18.5 & 2.3E-16 & 24.62 & 23.44 & 6.8 &G202,L265,X412,MY33403\\
M5522  & 53.157639 &  -27.849911 & 3.690 (2.7-4.1) & 0.3-3.0 &  3.0 & 2.4E-17 & 25.41 & 24.31 & 1.2 & MY33385    \\
M5842  & 53.070236 &  -27.845531 & 3.830 (2.8-4.3) & 0.7-1.8 &  3.5 & 3.2E-17 & 25.20 & 24.68 & 3.5 & X235, MY34075\\
M7305  & 53.072567 &  -27.827642 & 3.440 (3.1-3.7) & 0.9-1.8 &  3.5 & 3.5E-17 & 24.37 & 24.45 & 1.7 &            \\
M8039  & 53.034122 &  -27.817532 & 3.190 (2.5-3.9)$^d$& 0.5-5.0 &  3.1 & 1.7E-17 & 26.95 &-26.24 & 0.1 & MY38702    \\
M8273  & 53.165268 &  -27.814058 & 3.064 spec.     & 0.9-6.3 & 31.5 & 8.0E-16 & 24.51 & 22.40 &16.7 & MY39298,G27,L341,X546\\
M8728  & 53.069000 &  -27.807203 & 4.820 (3.9-5.5) & 0.8-2.2 &  3.1 & 5.4E-17 & 26.58 &-26.24 & 0.7 &            \\
M9350  & 53.167900 &  -27.797960 & 3.360 (3.2-3.5) & 0.4-3.6 &  3.1 & 1.7E-17 & 24.76 & 24.74 & 0.8 & MY41761    \\
M9363  & 53.051006 &  -27.797874 & 3.610 (3.0-4.1)$^d$& 1.3-4.0 &  3.2 & 8.4E-17 & 26.82 &-26.24 & 0.1 &            \\
M10390 & 53.059620 &  -27.784552 & 3.210 (2.4-3.7) & 0.3-3.6 &  2.8 & 2.0E-17 & 26.32 & 24.90 & 1.0 &            \\
M10429 & 53.178471 &  -27.784035 & 3.193 spec.     & 0.4-4.5 & 22.7 & 7.6E-16 & 25.18 & 24.08 & 2.0 & MY44068,L358,X573\\
M10515 & 53.143574 &  -27.783009 & 5.260 (1.8-6.6) & 1.0-6.0 &  3.3 & 7.3E-17 & 25.40 & 23.42 &21.1 &            \\
M10548 & 53.021168 &  -27.782366 & 4.823 spec.     & 0.4-1.9 &  3.2 & 4.4E-17 & 23.71 & 99.00 & 3.4 & MY44369    \\
M13473 & 53.171535 &  -27.743347 & 3.210 (2.6-3.8)$^d$& 0.6-1.8 &  3.1 & 2.6E-17 & 24.80 & 24.23 & 0.7 & MY50924    \\
M13549 & 53.020576 &  -27.742151 & 3.462 spec.     & 1.9-6.3 &  3.7 & 1.2E-16 & 23.69 & 23.33 & 3.5 & MY51260,E1577\\
M14078 & 53.117867 &  -27.734310 & 3.560 (2.8-3.8) & 0.3-4.0 &  5.2 & 4.7E-17 & 23.97 & 22.69 &19.9 & E2309,L252,X386 \\
M70091 & 53.158215 &  -27.733717 & 3.200 (2.0-5.4) & 0.9-2.0 &  4.2 & 4.2E-17 & 27.65 & 24.55 & 5.3 & E2551,G557,L323,X521\\
M70099 & 53.119797 &  -27.743099 & 6.220 ($>2.9$)  & 0.5-1.9 &  3.7 & 3.7E-17 & 27.10 & 24.91 & 6.2 & X392,E1516     \\
M70168 & 53.111530 &  -27.767839 & 5.010 (2.0-6.5) & 0.8-4.7 & 11.1 & 1.3E-16 & 25.97 &-25.03 & 2.3 & G518,L245,X371  \\
M70340 & 53.040970 &  -27.837717 & 3.900 ($>3.9$)  & 0.4-4.2 &  9.2 & 1.1E-16 &-27.56 &-25.46 & 2.2 & G537,L103$^e$, X156\\
M70385 & 53.107418 &  -27.855694 & 3.980 (1.1-4.7) & 0.6-2.0 &  7.3 & 1.2E-16 & 26.49 &-25.26 & 4.1 & G589,L235$^f$,X354\\
M70390 & 53.085411 &  -27.858023 & 4.180 (2.2-4.8) & 0.5-1.7 &  3.8 & 3.8E-17 & 26.20 & 23.98 & 5.7 & L191,X285   \\
M70407 & 53.064629 &  -27.862434 & 5.900 ($>$1.8)  & 0.9-1.9 &  2.8 & 2.8E-17 & 27.02 & 24.89 & 1.5 &            \\
M70429 & 53.137917 &  -27.868238 & 3.630 (1.2-4.8) & 0.4-3.4 &  9.7 & 1.5E-16 &-27.90 & 25.41 & 2.3 & G217a,L290$^g$,X456\\
M70435 & 53.215118 &  -27.870247 & 3.230 ($>$2.0)  & 0.7-6.0 &  9.8 & 1.2E-16 & 26.88 & 24.06 & 4.9 & G508,L404,X651  \\
M70437 & 53.146461 &  -27.870945 & 4.420 ($>$2.7)  & 1.4-5.3 &  7.7 & 3.9E-17 & 27.62 &-25.35 & 3.5 & L306$^h$,X485   \\
M70467 & 53.124233 &  -27.882565 & 4.250 ($>$2.0)  & 0.9-3.8 &  3.0 & 1.8E-17 & 26.85 &-25.50 & 3.9 &            \\
M70481 & 53.189320 &  -27.888466 & 4.080 (3.3-4.5) & 0.3-6.7 &  2.9 & 1.1E-16 & 26.75 & 24.02 & 8.0 &            \\
M70508 & 53.210999 &  -27.907394 & 3.180 (2.0-4.0) & 0.3-5.7 &  4.6 & 9.7E-17 & 26.24 & 99.00 & 7.5 &            \\
M70525 & 53.136761 &  -27.917326 & 3.280 (0.8-3.8) & 0.9-6.7 &  3.1 & 1.2E-16 &-27.87 & 24.26 & 3.7 &            \\
M70531 & 53.151512 &  -27.926792 & 3.260 ($>$1.9)  & 1.7-4.5 &  3.0 & 1.1E-16 & 26.07 & 23.27 &19.7 &            \\
\hline                                                  
M3607$^h$  & 53.215858 &  -27.876823 & 3.468  spec     & 1.0-3.8 &  2.9 & 8.5E-17 & 24.93 & 99.00 & 0.9 &            \\
M70107$^h$ & 53.079334 &  -27.741474 & 3.356  spec     & 0.8-4.7 &  3.4 & 2.1E-17 & 26.85 & 23.77 &12.8 & E1611      \\
\hline
216554$^{i,h}$ & 53.068958 & -27.684222 & $\gs6.0$ & 0.4-0.8 &  2.4 & 1.4E-5$^l$& 25.8& --    & --  & \\  
\hline
\end{tabular}
}

$^a$L=Luo08, M=GOODS-MUSIC, MY=GOODS-MUSYC Cardamone et al. 2010;X=Xue11,G=Giacconi et al. 2002; 
$^b$ Gilli et al. 2011; 
$^c$ Norman et al. 2002;
$^d$secondary solution at z$<$0.5;
$^e$ Luo08 z=2.45; Luo08$^f$ 2.71; 
$^g$ Luo08 z=2.36; $^h$ Luo08 z$>7.4$; $^h$ just below detection threshold;
$^i$ Bouwens et al. 2006; $^l$ count rate.
\end{table*}

\subsection{X-ray and optical spectroscopy}

\begin{table*}
\caption{\bf Multiwavelength properties of $z>3$ galaxies}
\begin{tabular}{lccccccccl}
\hline
ID   & $\Gamma$ & logL(2-10keV) & X-ray  &  F(1.4GHz) & SFR          & Morph.$^b$ & A(0.16) & A(0.16) & Optical  \\
     &  X-ray   & ergs/s        & obsc.$^a$& $\mu$Jy   & M$_\odot$/yr  &  H band   & SED     & X-ray   & spec. \\
\hline
E537  &  -0.15$\pm$0.2& 43.8 & CT & 19.6$\pm6.2$ & 730$\pm230$ & E & $<2.6$ & $>1000$  & \\
E737  &   1.6$\pm$1.0 & 43.1 &    & 15.0$\pm6.2$ & 360$\pm150$ & E & & & \\
E1312 &   1.2$\pm$0.7 & 43.7 &    & & & P & & & \\
E1516 &   3.0$_{-1.4}^{+2.0}$&42.8& & 17.8$\pm6.4$ & 660$\pm240$ & E & & & \\ 
E1577 &  -0.9$_{-1.4}^{+0.9}$&43.6& CT & & & P & $<1.1$ & $>930$ & Em.: Ly$\alpha$ narrow, CIV \\
      &                    &    &    & & &   &         &       & Abs.:  CIV broad \\
E1617 &   0.9$\pm$0.3 & 44.4 & HO & & & E & $<2.6$ & $>190$ & \\
E2199 &   1.6$\pm$1.3 & 42.9 &    & & & P & & & \\
E2309 &   1.7$\pm$0.7 & 42.9 &    & & & E & & & \\
E2498 &   0.9$\pm$0.5 & 43.9 &    & 16.3$\pm6.4$ & 1500$\pm590$ & E & & & \\
E2551 &   1.8$\pm$0.7 & 43.2 &    & 33.5$\pm6.4^g$ & 1870$\pm360$ & E & & & \\
E2658 &   0.3$\pm$1.4 & 42.8 &    & & & P & & & \\
E4956 &   2.2$\pm$0.9 & 43.9 &    & & & E & & & \\
E5165 &   1.8$\pm$0.8 & 43.2 &    & & & E & & & \\
E6257 &   1.5$\pm$1.5 & 42.8 &    & & & E & & & \\
E7911 &   1.5$\pm$1.0 & 43.6 &    & & & P & & & \\
E8479 &  -0.1$\pm$0.9 & 43.5 & CT & & & E & $<1.1$ & $>470$ & \\
E10247&   --          & 43.1 &    & & & E & & & \\
\hline
ID   &  $\Gamma$ & logL(2-10keV) & X-ray  &  F(1.4GHz) & SFR       & Morph.$^b$ & A(0.16) & A(0.16) & Optical  \\
     &   X-ray   & ergs/s        & obsc.$^a$ & $\mu$Jy   & M$_\odot$/yr & $z$ band   & SED     & X-ray   & spec. \\
\hline
M208$^c$&  0.1$\pm$0.6    & 44.0 & CT & 19.4$\pm$6.5   & 900$\pm300$ & P & $<2.0$ & $>780$ & Em: Ly$\alpha$ narrow, NV broad\\
M3320  &   1.12$\pm$0.06  & 44.8 & HO & & & E & $<3.5$ & $>120$ & Em.: Ly$\alpha$ narrow\\
M3323  &   1.6$\pm$0.21   & 43.8 &    & & & P &        &        & Em.: Ly$\alpha$ broad\\
M4302  &   1.33$\pm$0.08  & 44.4 & HO & 17.6$\pm6.1$   & 450$\pm160$ & P & $<1.8$ & $>110$ & Em.: Ly$\alpha$, Ly$\beta$, Ly$\gamma$, \\
       &                  &      &    & & &   &        &        & Abs: NV, CII, OI broad  \\
M4417$^d$& 1.8$\pm$1.0    & 42.5 &    & 15.3$\pm6.3$   & 360$\pm150$& E &        &        & Abs: CIV, SiII, SiIV,\\
         &                &      &    &            &            &   &        &        & CII, OI, SiII \\
M4835  &   0.08$\pm$0.24  & 44.0 & CT & $36.0\pm6.2^g$ & 960$\pm160$ & E &        &        & Em.: Ly$\alpha$, CIV narrow\\
M5390$^e$& 0.18$\pm$0.16  & 44.3 & CT & $19.4\pm6.1$   & 530$\pm170$ & P & $<1.1$ & $>820$ & Em.: Ly$\alpha$, CIV, OVI, \\
       &                  &      &    & & &   &        &        & NV, CIV, HeII narrow\\
M8273  &   0.29$\pm$0.10  & 44.8 & HO & $49.6\pm6.2^g$ & 900$\pm110$ & E & $<2.0$ & $>380$ & Em.: Ly$\alpha$, CIV narrow\\
M10429 &   1.40$\pm$0.10  & 44.0 &    & & & P &        &        & Em.: CIV broad \\
M10548 &   2.75$\pm$1.30  & 43.0 &    & & & E &        &        & Em.: Ly$\alpha$ narrow, noisy\\
M13549$^f$&-0.9$_{-1.4}^{+0.9}$&43.6&CT & & & P &        &        & Em.: Ly$\alpha$ narrow, CIV, \\
       &                  &      &    & & &   &        &        & Abs: CIV broad\\
\hline
\end{tabular}
$^a$ CT=Compton thick (N$_H\gs10^{24}$ cm$^{-2}$); 
HO=Highly obscured, Compton thin,($10^{23}\ls$N$_H\ls10^{24}$ cm$^{-2}$);
$^b$ Morphology, P=point like, E=Extended;
$^c$ Gilli et al. (2011); $^d$ Maiolino et al. 2008, [OII], [OIII] H$\beta$
$^e$ Norman et al. (2002); $^f$=E1577; $^g$VLA-CDFS DR2 flux
\end{table*}

\begin{figure*}[t!]
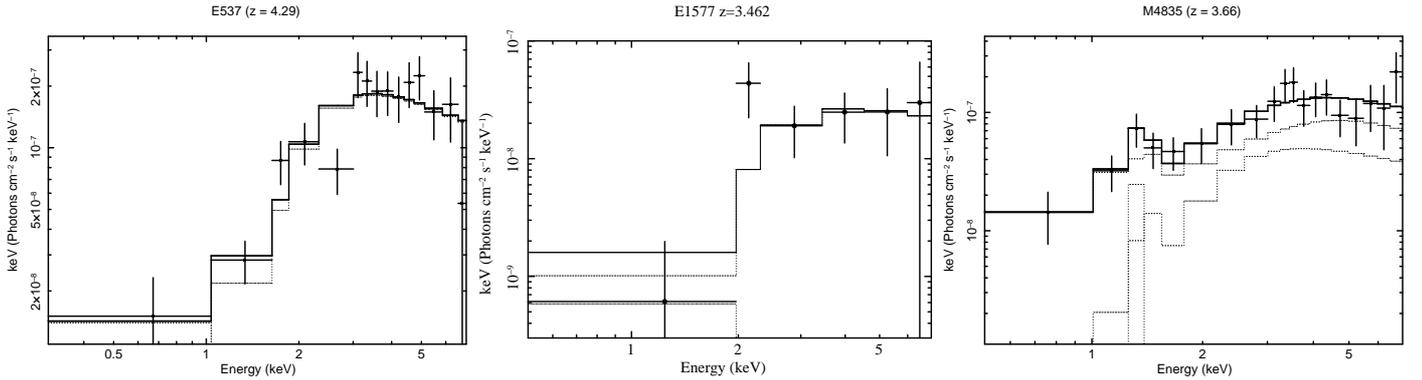

\centering
\begin{tabular}{cc}
\includegraphics[width=5cm,angle=-90]{f_e537_euf.ps}
\includegraphics[width=5cm,angle=-90]{f_e1577_euf.ps}
\includegraphics[width=5cm,angle=-90]{f_m4835_euf.ps}
\end{tabular}
\caption{The Chandra X-ray spectra of E537 [Left panel], 
E1577 [central panel], and M4835 [Right panel].}
\label{xspe}%
\end{figure*}

We discuss here X-ray and optical spectroscopy of our X-ray sources.
Unfortunately only one of the 17 sources in the Chandra-GOODS-ERS
sample has optical spectroscopy available. On the other hand, 11 of
the 41 sources of the Chandra-GOODS-MUSIC sample have optical
spectroscopy available.  To exploit this rather large dataset we
complement the full Chandra-GOODS-ERS sample, with the
Chandra-GOODS-MUSIC sources with optical spectroscopy\footnote{We do
  not discuss here the rest of the Chandra-GOODS-MUSIC sample because
  it is affected by incompleteness, see next section, and because the
  quality of the photometric redshift is worse than that of the
  GOODS-ERS sample}.

X-ray spectra have been extracted from the merged event file using
source extraction regions of 2-6 arcsec radius, depending on the
source off-axis (the Chandra Point Spread Function degrades quickly
with the off-axis angle). Background spectra have been extracted from
nearby source-free regions. We also used the total background spectra
at three off-axis angles (see the Appendix for details). Background
spectra are then normalized to the source extraction region area.  We
report in Table 3 the results of fitting a simple power law model to
the data.  Several Chandra-GOODS-ERS and Chandra-GOODS-MUSIC sources
have flat X-ray spectra, strongly suggesting the presence of large
absorbing column densities along those lines of sight. The quality of
the spectra is poor, however in several cases the statistic is good
enough to perform a proper fit with a model including an absorbed
power law or a reflection spectrum. In the following we briefly
outline the main results from this analysis.

\begin{itemize}

\item 
E537 shows an extremely hard spectrum. A simple power law fit produces
a negative slope of $\Gamma$ $\sim$ -0.15.  An absorbed power-law
model with $\Gamma$ fixed to 1.8 yields a column density of N$_H$=1.6
$\times$ 10$^{24}$ cm$^{-2}$. A better description of the spectral
shape of the low energy spectrum is obtained by adding an unabsorbed
power law (see Fig. \ref{xspe}).  Such component typically provides an
excellent fit of the X-ray spectrum of many well-studied obscured AGNs
and its origin is explained in terms of emission due to reprocessing
of the nuclear continuum by surrounding material and/or a portion of
primary radiation leaking through the absorber (see
e.g. \cite{turner:1997,piconcelli:2007}).  The photon index of the
soft X-ray component is fixed to $\Gamma$=1.8, while the resulting
ratio between the normalization of the unabsorbed and absorbed power
law is $\sim$0.02. An alternative model consisting of a Compton
reflection emission component ({\tt pexrav} model in XSPEC) obscured
by a column density of N$_H$ = 5$\pm$2 $\times$ 10$^{23}$ cm$^{-2}$ is
statistically equivalent.

\item 
The X-ray spectrum of M4835 reveals additional complexity when fitted
with a simple absorbed (N$_H$ = 9$\pm$2 $\times$ 10$^{23}$ cm$^{-2}$)
power-law model (C-stat/dof=425/442). The best-fit model to the {\it
  Chandra} data includes an additional component due to X-ray
reflection from cold circumnuclear matter. In this case the primary
continuum is obscured by a column density of
N$_H\sim1.5^{+0.4}_{-0.8}$ $\times$ 10$^{24}$ cm$^{-2}$ and the
normalization of the nuclear component is about 5 times larger than
that of the reflection component. Finally, the limit equivalent width
(EW) of the Fe K$\alpha$ emission line measured with respect to the
reflection continuum is $<$1.1 keV, i.e. still consistent with the
expectation of a reflection model.  Fig. \ref{xspe} shows the Chandra
spectrum of M4835 fitted by this reflection $+$ transmission model.

\item 
M5390 is G202, the well-known, highly obscured QSO discovered by
\cite{norman:2002} at $z$ = 3.7. The Chandra spectrum is consistent
with the XMM one presented by \cite{comastri:2011}.  It is well fitted
by a power law model (with $\Gamma=1.8$ fix) absorbed by a column
density of N$_H$ $\sim$ $1.2\pm0.15\times 10^{24}$ cm$^{-2}$ and
gaussian line (E=6.6 keV, likely a blend of the 6.4 keV neutral Fe and
6.7 keV Fe XXV K$\alpha$ lines, EW=0.8$\pm0.4$ keV), plus an
energetically unimportant soft component. The quality of the fit
decreases if the continuum is modeled with a pure reflection
component.

\item 
M208 is the source presented by \cite{gilli:2011}, and we confirm the
results reported in that paper, finding the spectrum consistent with
being obscured by a Compton thick absorber.  Very similar results are
obtained for E1577, which is best fitted by a heavily absorbed
(i.e. N$_H$ $\geq$ 10$^{24}$ cm$^{-2}$) power law model (see
Fig. \ref{xspe}).  We also find the presence of obscuration with
column density values consistent with 10$^{24}$ cm$^{-2}$ in the
spectrum of E8479.

\item 
Large N$_H$ values exceeding 10$^{23}$ cm$^{-2}$ (but still in the
Compton-thin regime) are reported from the spectral analysis of E1617,
M8273, M4302 and M3320. Remarkably, in the spectrum of the latter
source at z=3.471, we find clear evidence of an emission line with an
EW = 160$\pm$100 eV at a best-fit rest-frame energy of 6.97 keV and,
therefore, associated with highly ionized (i.e. Fe XXVI) iron.

\end{itemize}

In conclusion, at least 3 of the 17 Chandra-GOODS-ERS sources are
likely to be Compton thick AGN, which is $18^{+17}_{-10}\%$ of the AGN
in the sample are Compton thick, as well as 4 of the 11
Chandra-GOODS-MUSIC AGN with spectroscopic redshift (one of these
sources is in common with the Chandra-GOODS-ERS sample). Other 3
Chandra-GOODS-MUSIC and 1 Chandra-GOODS-ERS AGN are highly
obscured. The luminosity of the Compton thick ERS and GOODS-MUSIC
AGN is in the range log(L2-10keV)=43.5-44.5, i.e. between bright
Seyfert 2 galaxies and type 2 QSOs. The 2-10 keV flux of the
Chandra-GOODS-ERS sources is between 0.3 and 3$\times10^{-16}$\cgs. At
this fluxes the \cite{gilli:2007} model for the Cosmic X-ray
background (CXB) predicts a fraction of Compton thick AGN of
$\sim20$\%. This is already similar to the fraction of Compton thick
AGN we find at z$>3$, thus suggesting that Compton thick AGN are
probably more common at high-z than previously predicted (also see
\cite{gilli:2011}).

It is instructive to compare the result of the X-ray spectroscopy to
those of the optical spectroscopy, photometry and morphology in Table
3. The two CT AGN for which CIV is redshifted in the band covered by
optical spectroscopy show a narrow CIV, in addition to narrow
Ly$\alpha$. One of these (E1577/M13549) also shows broad absorption
blueward CIV.  The highest redshift CT AGN (M208) shows a narrow
Ly$\alpha$ and a relatively broad (2000km/s) NV emission (also see
\cite{gilli:2011}). Two of the three highly obscured AGN (M3320 and
M8273) also show narrow Ly$\alpha$ emission and, when redshifted in
the band covered by the spectroscopy, narrow CIV.  In the spectrum of
the third highly obscured source (M4302) strong and broad absorption
lines are present.

Four of the ten highly obscured and Compton thick AGN have a point like
morphology in the $z$ band (M13549/E1577, M5390, M4302, M208) or H
band (E1577) based on {\sc CLASS\_STAR} parameter of {\sc sextractor},
FWHM measurements, and eye-ball inspection of $z$ and H band
images. This suggests that the active nucleus contributes
significantly to the UV and optical rest-frame emission, as further
confirmed in M4302 and M13549/E1577 by the detection of likely nuclear
broad absorption lines in their UV rest frame spectra.

\begin{figure}
\centering
\includegraphics[width=8cm]{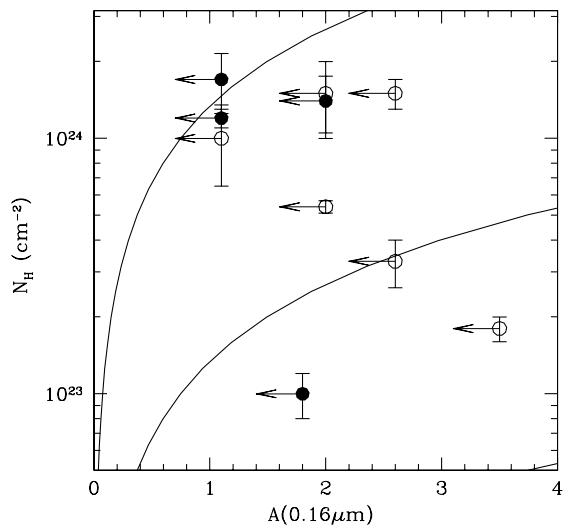}
\caption{ The best fit N$_H$ as a function of the 0.16$\mu$m
  extinction for the 10 z$>3$ highly obscured or Compton thick
  AGN. Filled points are sources with point like optical or near
  infrared morphology.  The two solid lines are the expectation for a
  dust-to-gas ratio 100 and 1000 times lower than the Galactic one.}
\label{nha16}%
\end{figure}

We estimated the rest-frame UV (0.16$\mu$m) extinction of the highly
obscured and Compton thick AGN by fitting their observed Spectral
Energy Distribution with galaxy templates and the \cite{calzetti:2000}
extinction law.  Table 3 gives the 1$\sigma$ upper limit to
A(0.16$\mu$m) along with the A(0.16$\mu$m) lower limit obtained using
the 1$\sigma$ lower limit to the neutral gas absorption column density
and assuming a Galactic dust-to-gas ratio.  We also plot for these
sources the best fit N$_H$ and A(0.16$\mu$m) upper limits in
Fig. \ref{nha16}.  We see that N$_H$ measurements and the
A(0.16$\mu$m) limits of all ten highly obscured or Compton thick AGN
are completely inconsistent with a Galactic dust-to-gas ratio. This is
true also for the four objects with point like morphology. The same
conclusion is reached by estimating dust extinction from the galaxy
stellar masses.  It is difficult to estimate accurate stellar masses
for our high-z AGN, but they likely exceed several $10^{10}$
M$_{\odot}$.  The extinction expected based on the correlations found
by \cite{pannella:2009} in a sample of z$\sim2$ BzK galaxies is of 3-6
magnitudes at 1500 \AA\ . We conclude that, at least for the four
point-like objects, the dust-to-gas ratio of the absorbing matter is
likely 1/100-1/1000 that of the Galaxy. This is also lower than in the
(mostly low-z) AGN studied by \cite{maiolino:2001}, and
\cite{shi:2006}.  The low dust-to-gas ratio may be explained if the
absorbing matter is within of close to the dust sublimation
radius. The other six highly obscured or Compton thick AGN have
extended morphologies in the $z$ and H images, suggesting that young
stellar populations significantly contribute to the UV and optical
rest frame emission. In these cases little can be said about the
nuclear extinction and the dust-to-gas ratio.

\subsection{Radio counterparts and star-formation rates}

Two Chandra-GOODS-ERS sources (E2551 and E1611/M70107) and other five
Chandra-GOODS-MUSIC sources (M2690, M4835, M8273, M70091 and M70340)
have a detection at 1.4 GHz in the DR2 catalog of the VLA-CDFS survey
\cite{miller:2008}. Radio fluxes for the Chandra-GOODS-ERS sources and the
Chandra-GOODS-MUSIC sources with spectroscopic redshift are given in Table 3.
Other 8 sources in Table 3 have a faint radio signal at the position
of the galaxy (signal to noise between 2.4 and 3.2). The probability
that this signal is a background fluctuations is $<2\%$, corresponding
to $<1$ spurious radio detection in the full sample of sources in
Table 3. We report in Table 3 radio fluxes also for the additional 8
faint detections.  We stress that the latter faint fluxes may
over-estimate the real radio flux because of the Eddington bias. A
more robust average flux can be obtained by stacking together the
radio images at the position of the X-ray sources.  The average flux
for the Chandra-GOODS-ERS and Chandra-GOODS-MUSIC sources without a
detection in the DR2 VLA-CDFS catalog is 6.5$\pm$1.5$\mu$Jy and
10.1$\pm$1.1$\mu$Jy respectively.

The observed radio fluxes can be due to both nuclear and stellar
processes. Using the \cite{lafranca:2010} probability distributions
for the X-ray to radio nuclear luminosity we would expect a
L(2-10keV)/L(1.4 GHz) ratio $<3\times10^4$ for $\sim25\%$ of the
sources in Table 3, and L(2-10keV)/L(1.4 GHz)$<10^3$ for 5\% of the
sources.  We converted the radio fluxes in Table 3 into luminosities by 
using the following expression:

\begin{equation}
L(1.4{\rm GHz})=1.19\times 10^{20} DL^2 f(1.4 {\rm GHz}) 
(1+z)^{(\alpha-1)}
\end{equation}

where $DL$ is the luminosity distance in Mpc and $f(1.4 {\rm GHz})$ is
in $\mu$Jy and we used $\alpha=0.7$. We find in all cases
L(2-10keV)/L(1.4 GHz)$<3\times10^4$, which is 41\% of the full sample,
and L(2-10keV)/L(1.4 GHz)$<10^3$ for 5 sources, i.e. 18\% of the full
sample.  This difference between the expected and observed fraction of
radio bright sources is not conclusive, because our radio fluxes may
be over-estimated as discussed above.  However, if we associate to the
sources without a detection in the DR2 VLA-CDFS catalog the average
flux measured above, the fractions of sources with L(2-10keV)/L(1.4
GHz)$<3\times10^4$ and L(2-10keV)/L(1.4 GHz)$<10^3$ are 92\% and 19\%
respectively.  We can therefore conclude that, at least on average,
radio fluxes are higher than expected based on the X-ray fluxes and
assuming common nuclear origin.  If radio emission is dominated by
stellar processes we can convert radio fluxes in star-formation rates
(SFRs, see e.g. \cite{yun:2001}). We used the following calibration
which assumes a Chabrier IMF\footnote{assuming a Salpeter IMF would result in 
SFR $\sim1.7$ times higher}:

\begin{equation}
SFR = 3.4 \times 10^{-22}L(1.4{\rm GHz}) 
\end{equation}

In two cases we have independent estimates of SFR, which agrees
reasonably well with the values in Table 3.  M208, has a SFR$\sim
1000$M$_\odot$/yr estimated from a Laboca 870$\mu$m detection (see
\cite{gilli:2011}), while M4417 has SFR$\sim440$M$_\odot$/yr,
estimated from UV SED fitting and oxygen lines (\cite{maiolino:2008}).

\subsubsection{A high-z star-forming galaxy?}

\begin{figure*}
\centering
\includegraphics[width=17cm]{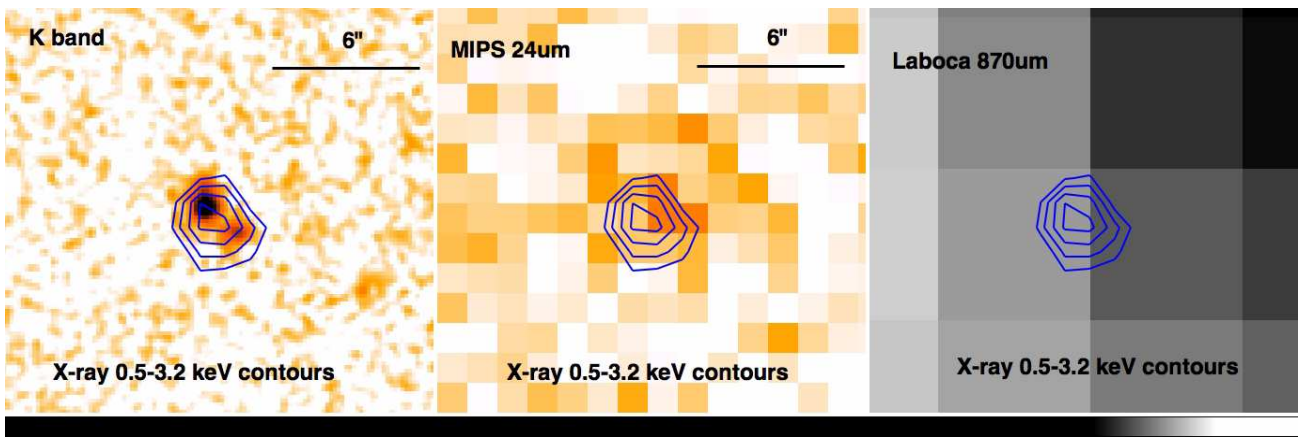}
\caption{ K band, MIPS 24$\mu$m and LABOCA 870$\mu$m images around the
  M4417 galaxy with overlaid X-ray 0.5-3.2 keV contours.}
\label{4417}%
\end{figure*}

Rather peculiar is the case of M4417, a galaxy studied by
\cite{maiolino:2008} in the framework of the AMAZE program.  The X-ray
luminosity corresponding to the observed SFR is
logL(2-10keV)$\sim42.35$ (by using the \cite{ranalli:2003} conversion
factor), just a factor 40\% lower than the observed X-ray
luminosity. It should be noted that although the Chandra X-ray
contours seems centered on M4417 (see Fig. \ref{4417}), there could be
some contribution to the observed X-ray flux from the nearby z=3.471
galaxy M4414, which has a SFR $\sim1/4$ that of M4417.  Intriguingly,
M4417 is the Chandra-GOODS-MUSIC and Chandra-GOODS-ERS object with the
lowest X-ray to $z$ band and H band flux ratios (0.004 and 0.005
respectively), lower than typical AGN (see next section), further
suggesting that the X-ray emission from this object is not dominated
by a nuclear source. We then conclude that the observed X-ray
luminosity is likely dominated by stellar sources, making M4417 one of
the farthest objects in which X-rays are probing stellar processes. Of
course other solutions are possible, for example M4417 may be a
reflection dominated, heavily Compton thick AGN, where the nucleus is
completely hidden and the observed X-ray luminosity is just a fraction
of the real one. We do not have however evidences in this direction
from either X-ray colors, optical spectroscopy, infrared spectroscopy
and broad band UV to infrared SED (the source is not detected at
wavelengths longer than 8$\mu$m, see Fig. \ref{4417}).

There are other two sources with relatively low X-ray luminosity and
high radio flux: E2551 and E1516. In these cases the X-ray luminosity
corresponding to the SFR rate estimated from the radio flux is a
factor of 70\% and 90\% lower than the observed X-ray luminosity.
Their X-ray to H band ratio are however $\gs10$ times higher than that
of M4417 (0.06 for E2551 and 0.07 for E1516). The radio flux and
luminosity of E2551 are the highest in the sample, suggesting that
this may well be a radio loud AGN.

\subsection{X-ray to optical/NIR flux ratios and selection effects}

\begin{figure*}[t!]
\centering
\begin{tabular}{cc}
\includegraphics[width=8.5cm]{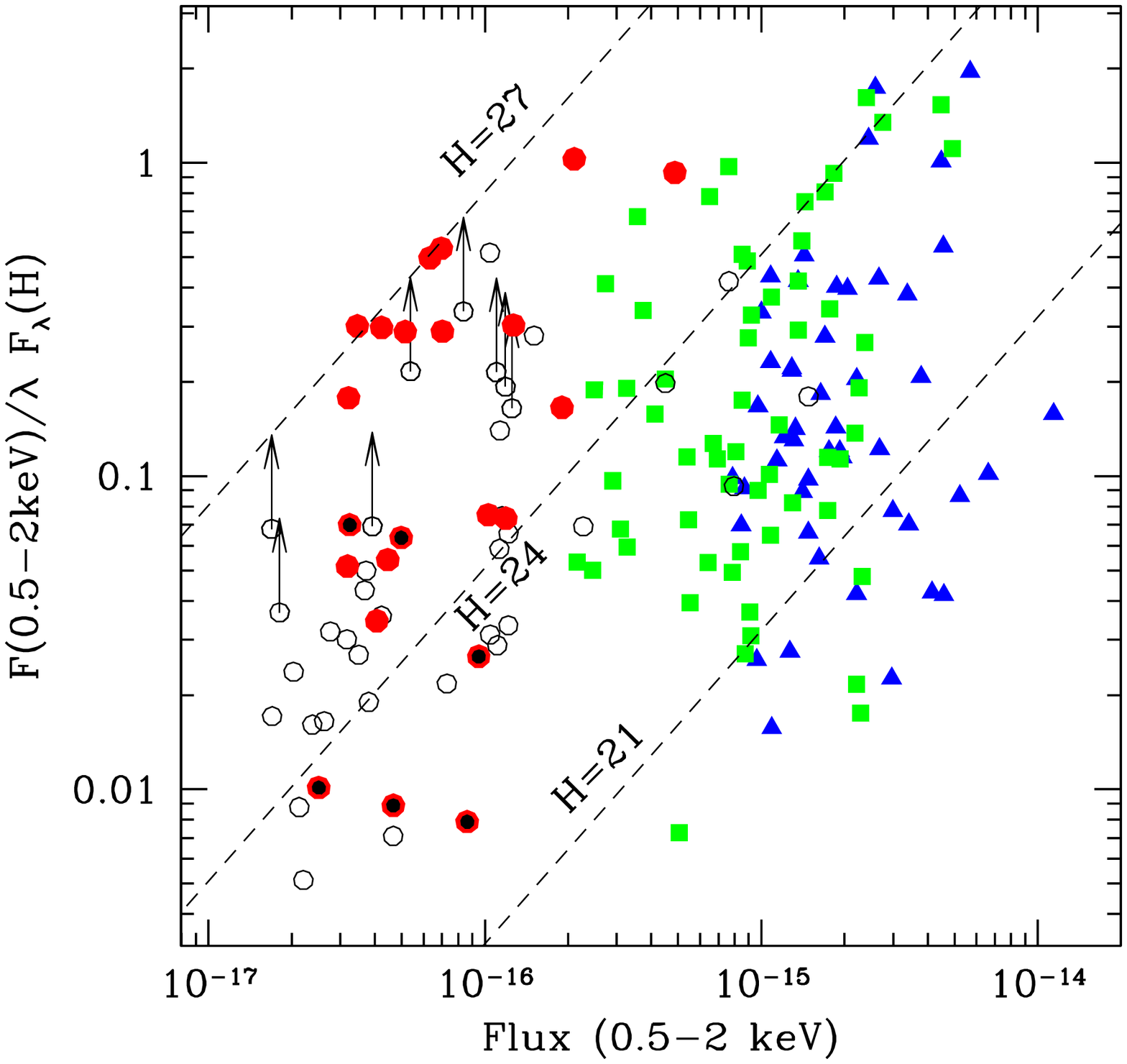}
\includegraphics[width=8.5cm]{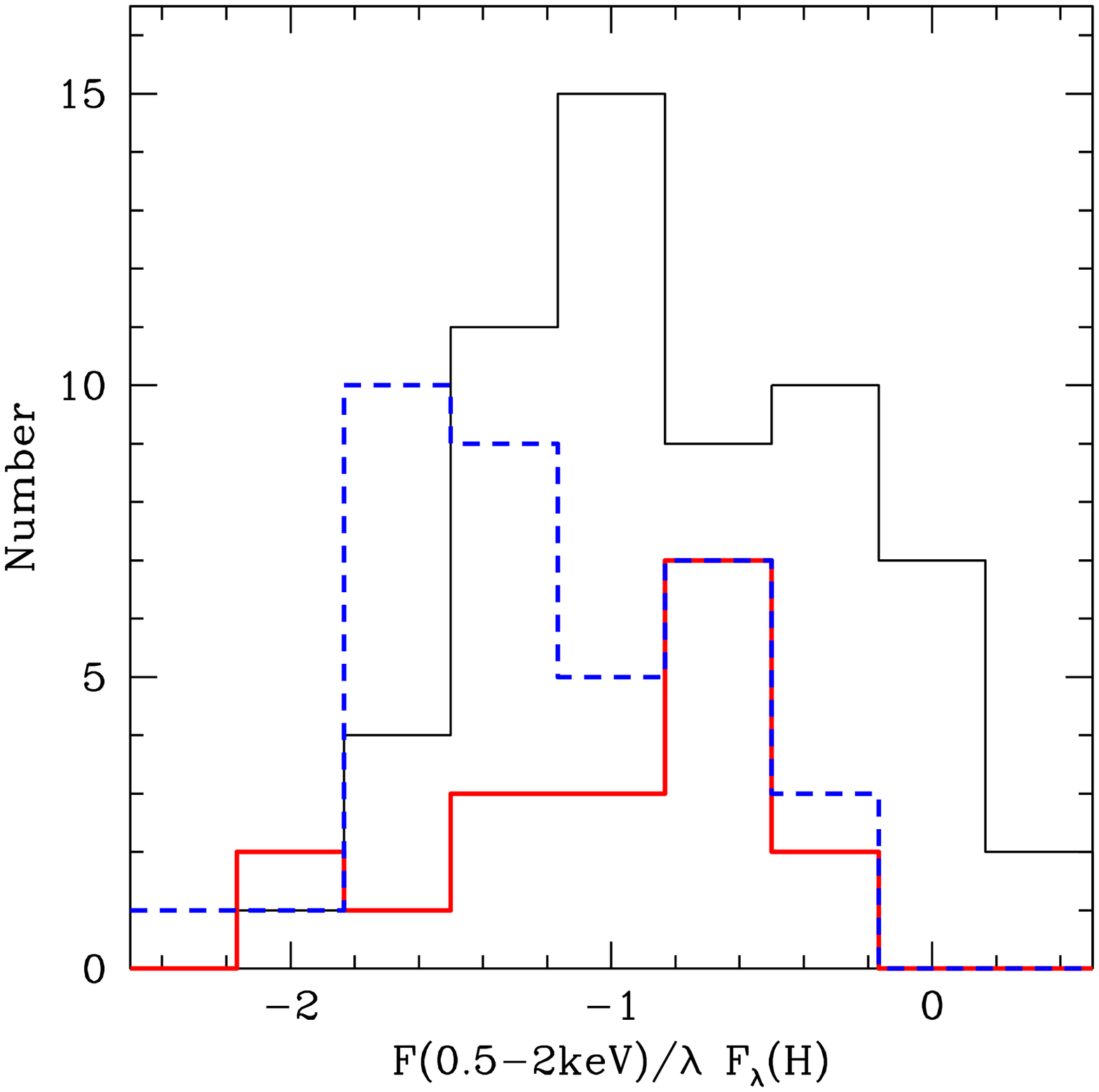}
\end{tabular}
\caption{[Left panel]: The X-ray to NIR flux ratio F(0.5-2keV)/F(H) as
  a function of the X-ray flux for the z$>3$ Chandra-GOODS-ERS AGN sample
  (red-filled circles), Chandra-GOODS-MUSIC sample (black-open
  circles), Chandra-COSMOS sample (green-filled squares) and
  XMM-COSMOS sample (blue-filled triangles. [Right panel:] The
  distributions of F(0.5-2keV)/F(H) for the Chandra-COSMOS z$>3$
  sample (black histogram), Chandra-ERSVO sample (red histogram), and
  Chandra-GOODS-MUSIC sample (blue, dashed histogram).}
\label{xir}%
\end{figure*}

The standard procedure to assemble X-ray high-z AGN samples is through
the optical identification of X-ray source catalogs. We used in the
previous section a complementary approach, that is studying the X-ray
emission of optically or NIR selected high-z galaxies. In the first
case the problem is to be sure that the optical/NIR identification of
the X-ray source is correct, and that sample identifications are
reasonably complete. In the latter case the problem is the reliability
and completeness of the optical/NIR samples. They should include most
of the galaxies (and therefore most AGN) down to a reasonably low flux
limit. The GOODS-MUSIC catalog is fairly complete down to $z\sim26$
and F(4.5$\mu m$)$\sim1.5 \mu$Jy and includes sources down to
$z\sim27$ and F(4.5$\mu m$)$\sim0.5 \mu$Jy. The GOODS-ERS catalog is
fairly complete down to H$\sim26.5$ and includes sources down to
H$\sim27.5$. To understand how these optical and NIR flux limits
translate into X-ray flux limits we compute the X-ray (0.5-2keV) to
NIR (H band) flux ratio F(0.5-2keV)/F(H) of much brighter, and
therefore likely complete, AGN samples.  Fig. \ref{xir} compares
F(0.5-2keV)/F(H) of the z$>3$ Chandra-GOODS-ERS and Chandra-GOODS-MUSIC AGN
samples to the Chandra-COSMOS (\cite{civano:2011}) and XMM-COSMOS
(\cite{brusa:2009a,brusa:2010}) z$>3$ AGN samples. The COSMOS samples
are X-ray selected samples with nearly complete optical
identification, thanks to the fact that they cover X-ray and
optical/NIR flux ranges 10-100 times brighter than the Chandra-GOODS-ERS
and Chandra-GOODS-MUSIC AGN samples, and the massive photometric
campaigns performed in this field with HST/ACS, Subaru and CFHT (Capak
et al. 2008). As an example, there are only two sources in the
Chandra-COSMOS catalog without an optical or infrared counterpart, 19
without an optical counterpart, 0.1\% and 1\% of all Chandra-COSMOS
sources respectively, \cite{civano:2011}. Most of these sources are
likely high luminosity highly obscured type 2 QSOs
(e.g. \cite{fiore:2003}).  But even assuming that {\it all} these 19
sources are at high-z, they would be 25\% of the z$>3$ sources in
Chandra-COSMOS. So this is a very conservative upper limit to the
Chandra-COSMOS completeness at high-z. Similar numbers apply to the
XMM-COSMOS survey.

The Chandra-GOODS-ERS detections cover a range of F(0.5-2keV)/F(H) similar
to the COSMOS samples and indeed the Chandra-GOODS-ERS and COSMOS
distributions of F(0.5-2keV)/F(H) are perfectly consistent (the
probability that the Chandra-GOODS-ERS and Chandra-COSMOS distributions
are drawn from the same parent population is 50\%, using the
Kolmogorov-Smirnov test).  This suggest that the HST/WFC3 ERS H band
images are deep enough to trace high-z AGN populations at the
extremely faint X-ray flux limits reached by the Chandra 4 Msec
exposure, avoiding significant incompleteness, or at least with an
incompleteness comparable to that reached at much brighter fluxes by
Chandra-COSMOS and XMM-COSMOS.  The same exercise performed using the
GOODS-MUSIC $z$ band and IRAC 4.5$\mu m$ selected catalog is not
satisfactorily (we used to this purpose a subsample of the full
Chandra-GOODS-MUSIC sample with deep H band VLT/ISAAC coverage, 
excluding 13 sources with shallower H band coverage).  The
Chandra-GOODS-MUSIC sample clearly misses sources with high
F(0.5-2keV)/F(H) flux ratio in comparison to the Chandra-GOODS-ERS and
COSMOS samples, and indeed the Chandra-GOODS-MUSIC F(0.5-2keV)/F(H)
distribution is inconsistent with the Chandra-COSMOS distribution at
the 99.925 \% confidence level (using the Kolmogorov Smirnov test).
This means that the GOODS ACS and IRAC images are not deep enough to
fully probe the AGN population at the flux limits of the CDFS 4 Msec
observation. The F(0.5-2keV)/F(H) distribution of the
Chandra-GOODS-MUSIC sources with F(0.5-2keV)$\gs5\times10^{-17}$ \cgs
is consistent with the GOODS-ERS and COSMOS ones (probability of
$\sim30\%$ that they can be drawn from the same parent population)
and therefore in the following we will consider only the 23
Chandra-GOODS-MUSIC sources brighter than this flux limit (and outside
the ERS area). These 23 sources are added to the 17 sources of
Chandra-GOODS-ERS sample to form a total sample of 40 high-z AGN. We use
this sample to constrain the faint end of the AGN luminosity function
at high redshift.

\section{High-z AGN luminosity functions}

\begin{figure}
\centering
\includegraphics[width=8cm]{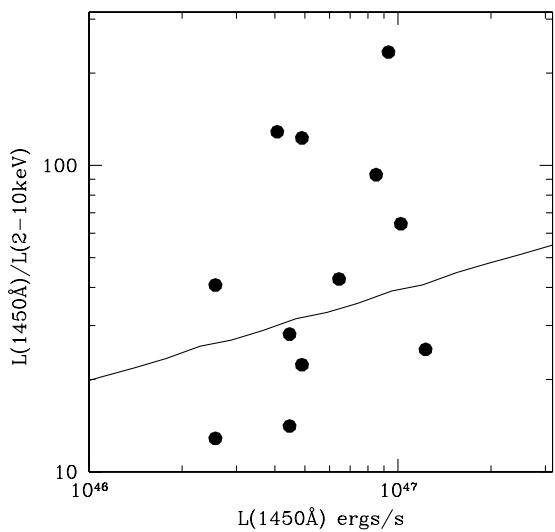}
\caption{The  UV 1450\AA\ to X-ray 2-10 keV luminosity ratio
for 12 QSOs at z$>5.7$}
\label{sdssqso}%
\end{figure}

\begin{figure}
\centering
\includegraphics[width=7cm]{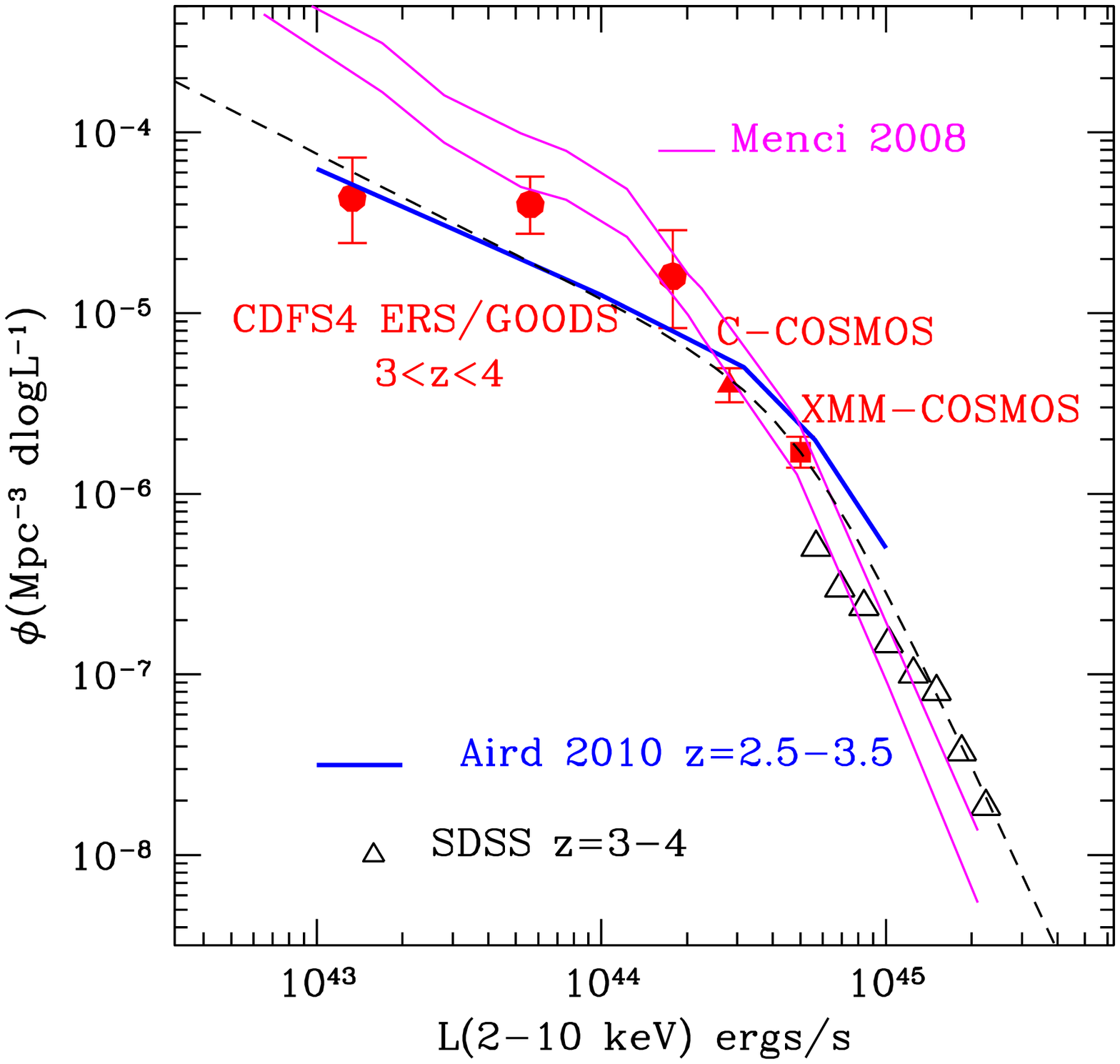}
\includegraphics[width=7cm]{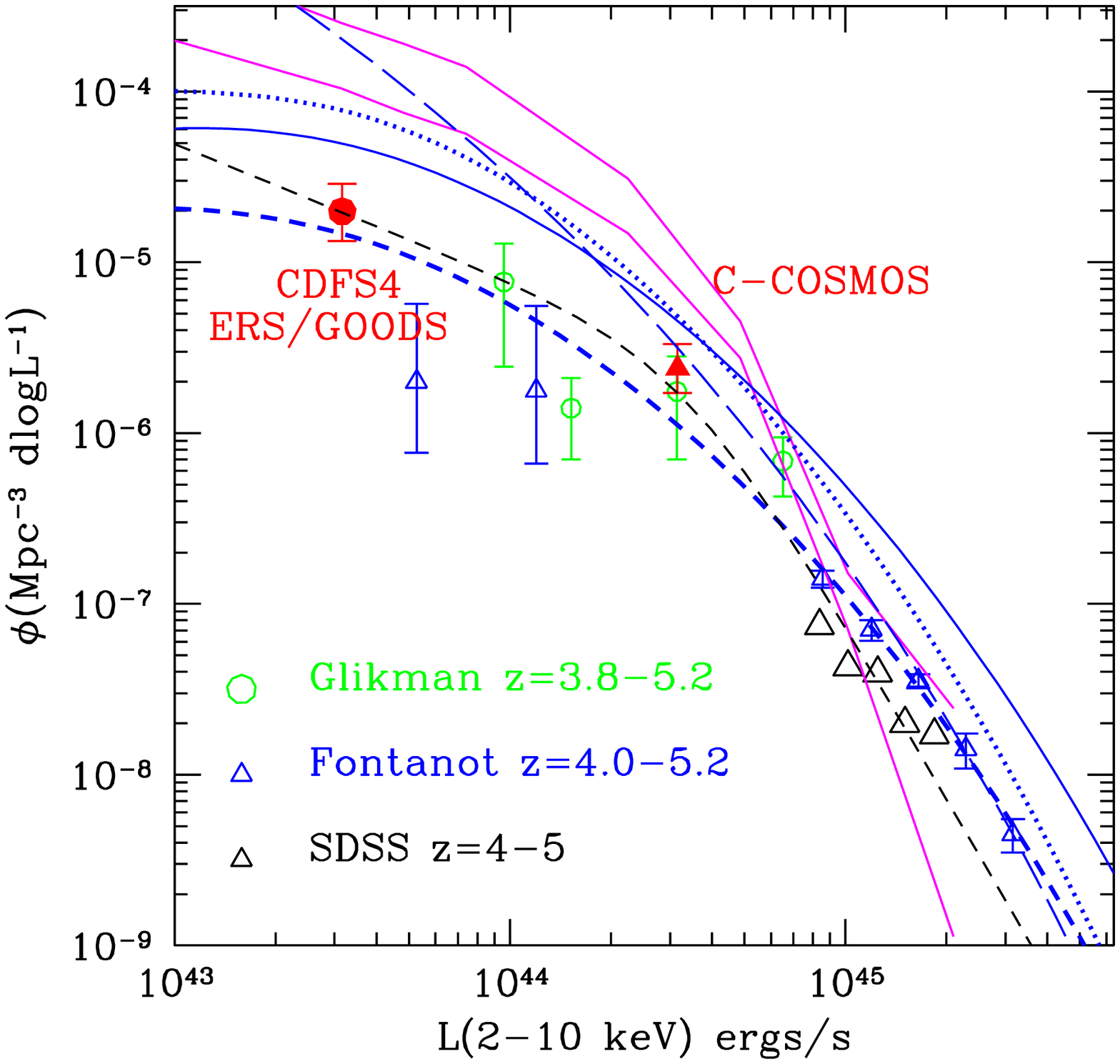}
\includegraphics[width=7cm]{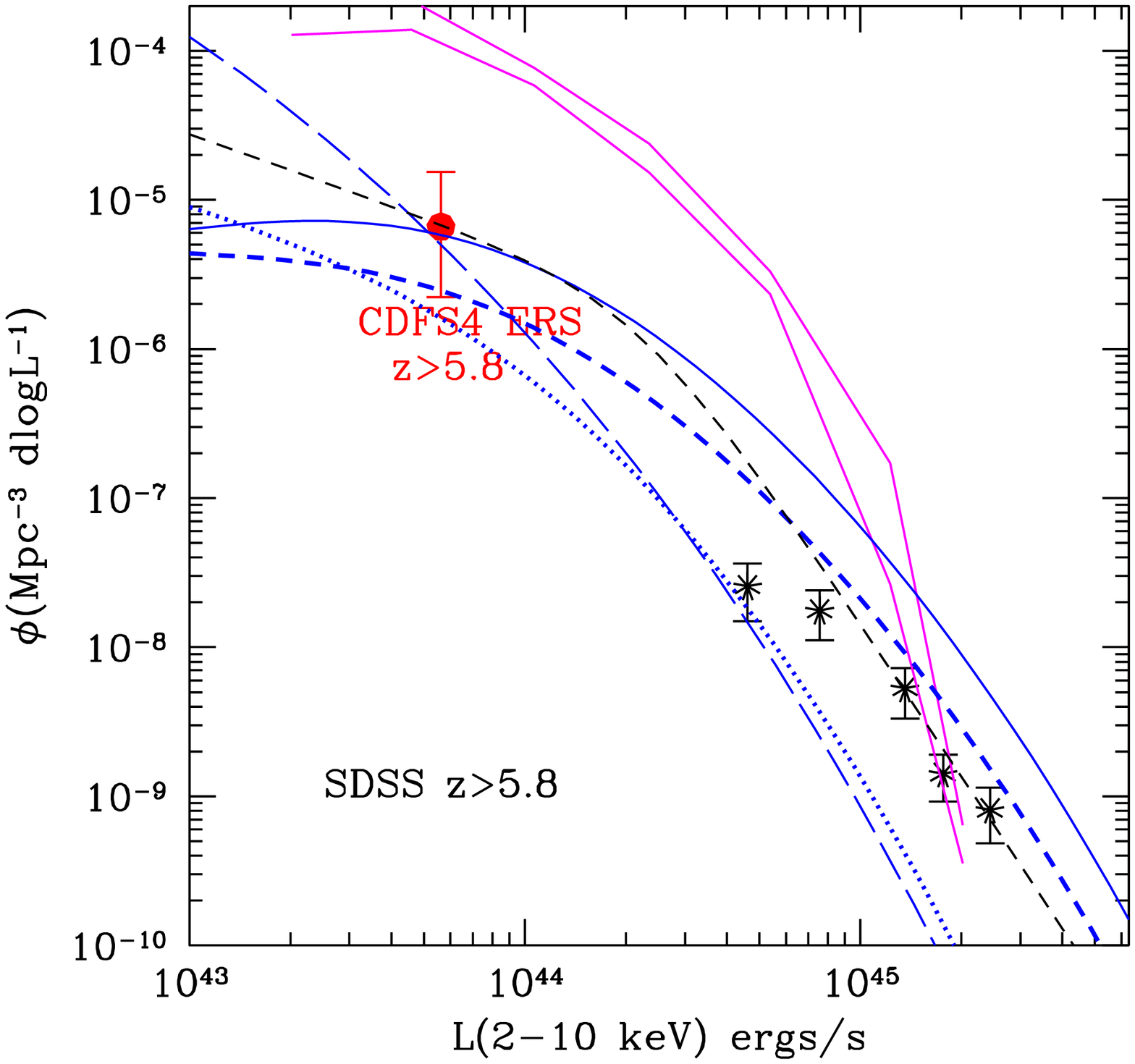}
\caption{The AGN 2-10 keV luminosity functions at z=3-4 (top panel),
  z=4-5 (central panel) and z$>$5.8 (bottom panel).  Large red-filled
  circles are from the CDFS Chandra-GOODS-MUSIC and Chandra-GOODS-ERS
  samples. Red-filled triangles are from Chandra-COSMOS (Civano et
  al. 2011), red-filled squares from XMM-COSMOS (Brusa et al. 2010).
  The blue thick curve in the top panel is the LADE model of Aird et
  al. 2010 in the 2.5-3.5 redshift bin. Large black-open triangles are
  SDSS data from Richards et al. 2006, stars are SDSS data from Jiang
  et al. 2009, open circles are from Glikman et al. 2011 and small
  blue-open triangles are from Fontanot et al. 2007. Black dashed
  curves are best fits to the X-ray plus optical, wide luminosity
  range luminosity functions (see text for details). Magenta solid
  curves encompass predictions on the Menci et al. (2006, 2008)
  semi-analytic model. Blue curves are models from Shankar et
  al. (2011), see the Discussion for details.  }
\label{hzlf}%
\end{figure}

Our z$>3$ AGN candidate sample includes 17 Chandra-GOODS-ERS sources
and 23 Chandra-GOODS-MUSIC sources spanning a range in luminosity
42.5$<$logL(2-10keV)/ergs/s$<$44.8. They can be used to probe the
faint end of the high-z AGN luminosity functions.  We calculated the
comoving space densities of our high-z sample using the $1/V_{max}$
method (\cite{schmidt:1968}). We choose L(2-10keV) and z ranges to
ensure completeness at the X-ray flux limit
F(0.5-2keV)$\sim2\times10^{-17}$ \cgs reached by our survey.  The
luminosity limit at z=4, 5 and 7.5 is
logL(2-10keV)$\sim42.7,~42.8,~43.3$ respectively.  Accordingly we
compute comoving space density in the redshift bins 3-4, 4-5 and
5.8-7.5 with luminosity ranges 42.75-44.5, 43-44 and 43.5-44.5
respectively. There are a total of 30 Chandra-GOODS-ERS and
Chandra-GOODS-MUSIC AGN in these redshift and luminosity bins.
Comoving space densities are given in Table 4. Fig. \ref{hzlf}
presents the AGN luminosity functions in the same redshift
bins. Errors are computed using Poisson statistics on the number of
AGN in each redshift-luminosity bin. To obtain information on the
slope of the luminosity functions we joined the above samples with the
Chandra-COSMOS high-z sample of \cite{civano:2011}, the XMM-COSMOS
sample of \cite{brusa:2009a}, the GOODS sample of
\cite{fontanot:2007}, the faint optical AGN sample of
\cite{glikman:2011}, the luminous optical AGN samples of
\cite{richards:2006} and \cite{jiang:2009}.  We converted the rest
frame 1450 \AA\ luminosities of the optically selected samples to the
2-10 keV band using the \cite{marconi:2004} and Sirigu et al. 2011
luminosity dependent conversion factors. We assume no intrinsic
reddening in the optically selected, high-z AGN. This might
underestimate the real AGN luminosity, if significant dust is present
along the high-z AGN lines of sight
(\cite{maiolino:2004,jiang:2006,gallerani:2010}).  We checked our
conversion factors against real data, using the Chandra and XMM
detections of the \cite{jiang:2009} QSOs
(\cite{mathur:2002,shemmer:2006} and references therein).
Fig. \ref{sdssqso} shows the ratio between the UV 1450\AA\ luminosity
and the 2-10 keV luminosity against 1450\AA\ luminosity for 12 QSOs at
z$>5.7$ for which we collected X-ray data from the literature. The
figure also shows the UV to X-ray conversion adopted in this work,
which agrees quite well with the data of this high redshift QSO
sample.

\begin{table*}
\caption{\bf Chandra-GOODS-ERS z$>$3 AGN comoving space densities}
\begin{tabular}{lcccc}
\hline
logL(2-10keV) & 3-4                     &  4-5                   & 3.8-5.2                 & 5.8-7.5 \\
\hline
42.75-43.5 & $4.3^{+2.9}_{-1.9}\times10^{-5}$ &        --              &      --                  & \\ 
43.5-44.0  & $4.0^{+1.7}_{-1.2}\times10^{-5}$ &        --              &    --                    & \\ 
44.0-44.5  & $1.6^{+1.3}_{-0.8}\times10^{-5}$ &        --              &    --                    & \\ 
43-44      &        --              & 2.1$^{+0.9}_{-0.7} \times10^{-5}$ & 3.5$^{+1.9}_{-1.3} \times10^{-5}$  & \\
43.5-44.5  &        --      &           --       &          --        & 0.66$^{+1.1}_{-0.5} \times10^{-5}$ \\
\hline
\end{tabular}

\end{table*}

We modeled the wide luminosity range high-redshift luminosity
functions using the standard double power law shape:

\begin{equation}
\frac{{\rm d} \Phi (L_{\rm X})}{{\rm d Log} L_{\rm X}} 
= A [(L_{\rm X}/L_{*})^{\gamma 1} + (L_{\rm X}/L_{*})^{\gamma 2}]^{-1}.
\end{equation}

where $A$ is the normalization factor for the AGN density, $\gamma 1$
and $\gamma 2$ are the faint-end and bright-end slopes and $L_*$ is
the characteristic break luminosity.  Optical selection may miss
highly obscured AGN, which can represent a large fraction of the total
in particular at low luminosity (\cite{lafranca:2005}). To avoid
possible incompleteness in optical surveys we excluded from the fit
the optically selected AGN density determinations with
$M_{1450}>-26.5$ (or correspondingly L(2-10keV)$<$45.1).  We first
fitted this simple model to the luminosity functions in the three
redshift bins. The number of points in the z=3-4 redshift bin allows
us to constrain all four parameters. This is not possible in the
redshift bins 4-5 and $>5.8$, where we fitted the data fixing the
slopes $\gamma 1$ and $\gamma 2$ to the best fit values found in the
z=3-4 redshift bin.  The results of these fits are reported in Table
5. As a next step, we fitted simultaneously the data in the three
redshift bins with an evolutionary model.  We chose to describe the
evolution of the high-z AGN luminosity function with the {\it LADE}
(luminosity and density evolution) model, introduced by
\cite{aird:2010}.  Thus the evolution of $L_*$ is given by:

\begin{equation}
{\rm Log}L_*(z)={\rm Log}L_0+p{\rm Log}(1+z)
\end{equation}

and the evolution of  $A$ is given by:

\begin{equation}
{\rm Log}A(z)={\rm Log}A_0+d{\rm Log}(1+z)
\end{equation}

We first fitted the data with the full six parameter model. The
$\gamma 1$ and $\gamma 2$ slopes turned out similar to those obtained
fitting the z=3-4 data only (the z=3-4 data provides the strongest
constrain to the shape of the luminosity function). We report in Table
5 the best fit parameters obtained by fixing $\gamma 2$ to 3.4 and to
3.0 (keeping the faint-end slope $\gamma 1$ fixed at 0.8 in both
cases). The fit with $\gamma 2=3$ produces a $\chi^2$ significantly
worse than for the best fit case, but still acceptable considering the
uncertainties on the correction between the optical and X-ray
luminosities for the optically selected AGN density determinations.
We found that the $A_0$ and $d$ parameter are completely
degenerate. We could obtain equally good fits with combinations of
$A_0$ and $d$ inversely correlated. We report in Table 5 the best fits
obtained with $d=0$ (no density evolution) and $d=-1$. On the other
hand, the data are good enough to constrain relatively well $L_0$ and
$p$. Our data are thus consistent with a pure luminosity evolution,
with $L_*$ rather quickly reducing with the redshift.  The dashed
lines on Fig. \ref{hzlf} are the best fit pure luminosity evolution
model (4) in Table 5.

It is not straightforward to compare our high-z, wide luminosity band
luminosity functions with previous determinations. The best fit model
of \cite{shankar:2009a} agrees quite well with our determinations at
logL(2-10keV)$<44$ but overestimate the density of higher luminosity
AGN, in particular of optically selected AGN. It should be noted that
that \cite{shankar:2009a} adopt UV to X-ray luminosity conversion
factor fixed at 10.4, while our conversion factor varies with the
luminosity, from 4.4 at logL(2-10keV)=43 to 27 at
logL(2-10keV)=45. Furthermore, \cite{shankar:2009a} correct for
extinction and the fraction of Compton thick AGN. As discussed above
our data are not corrected for the fraction of Compton thick object,
because several have actually been found in our samples, and we use
optically selected AGN of high luminosity ($M_{1450}>-26.5$,
L(2-10keV)$<$45.1), where the fraction of obscured AGN is likely to be
small. \cite{hopkins:2007} use a luminosity dependent bolometric
correction, but with a different calibration with respect to
ours. They correct their data for extinction and the fraction of
Compton thick AGN missed in optical, X-ray and infrared surveys. They
convert N$_H$ distributions evaluated from X-ray data into optical-UV
reddening using a canonical gas-to-dust ratio, which, as discussed in
the previous section (and by e.g. \cite{maiolino:2001,shi:2006}), can
overestimate the real extinction, in particular when the X-ray
absorber is compact, smaller than the dust sublimation radius. They
find z$>3$ luminosity functions with much flatter slopes with respect
to ours (or to the \cite{shankar:2009a} ones). As an example, the
  \cite{hopkins:2007} best fit model would predict $\sim2-4$ times
  more z=3-4 and z=4-5 AGN with logL(2-10keV)$>44.5$ than actually
  found in the XMM and Chandra COSMOS surveys
  (\cite{brusa:2009a,civano:2011}).

\begin{table*}
\caption{\bf Modeling the high redshift AGN luminosity function}
\begin{tabular}{lcccccccc}
\hline
model  & z range & $A$ or $A_0$ & $L_*$ or $L_0$ & $\gamma 1$ & $\gamma 2$ & $p$ & $d$ & $\chi^2$ (dof)\\
       &         & $10^{-6}$ Mpc$^{-3}$ & $10^{44}$ ergs/s&     &            &     &     &               \\
\hline
1 & 3-4 & $3.0^{+4.9}_{-1.5}$ & $5.4\pm2.2$ & $0.8\pm0.2$ & $3.4^{+0.6}_{-0.4}$ & -- & -- & 4.63 (4) \\
2 & 4-5 & $3.2^{+1.2}_{-1.1}$ & $3.5\pm0.4$ & 0.8 FIX & 3.4 FIX & -- & -- & 2.52 (2) \\ 
3 & $>5.8$ & $2.4^{+3.3}_{-2.0}$ & $2.1^{+3.1}_{-0.5}$ & 0.8 FIX & 3.4 FIX & -- & -- & 1.23 (1) \\ 
\hline
4 & $>3$ & $3.3^{+0.7}_{-1.0}$ & $120^{+60}_{-30}$ & 0.8 FIX & 3.4 FIX & $-2.1\pm0.2$ & 0 FIX & 10.65 (14) \\ 
5 & $>3$ & $5.2^{+1.1}_{-1.6}$ & $130^{+70}_{-40}$ & 0.8 FIX & 3.0 FIX & $-2.4\pm0.2$ & 0 FIX & 16.09 (14) \\ 
6 & $>3$ & $1.6^{+0.3}_{-0.5}$ & $70^{+30}_{-20}$ & 0.8 FIX & 3.4 FIX & $-1.8\pm0.2$ & -1 FIX & 10.82 (14) \\ 
7 & $>3$ & $2.6^{+0.6}_{-0.8}$ & $75^{+40}_{-25}$ & 0.8 FIX & 3.0 FIX & $-2.03\pm0.15$ & -1 FIX & 15.77 (14) \\ 
\hline
\end{tabular}

\end{table*}

\subsection{AGN luminosity function and duty cycle evolution}

\begin{figure}[h!]
\centering
\includegraphics[width=7cm]{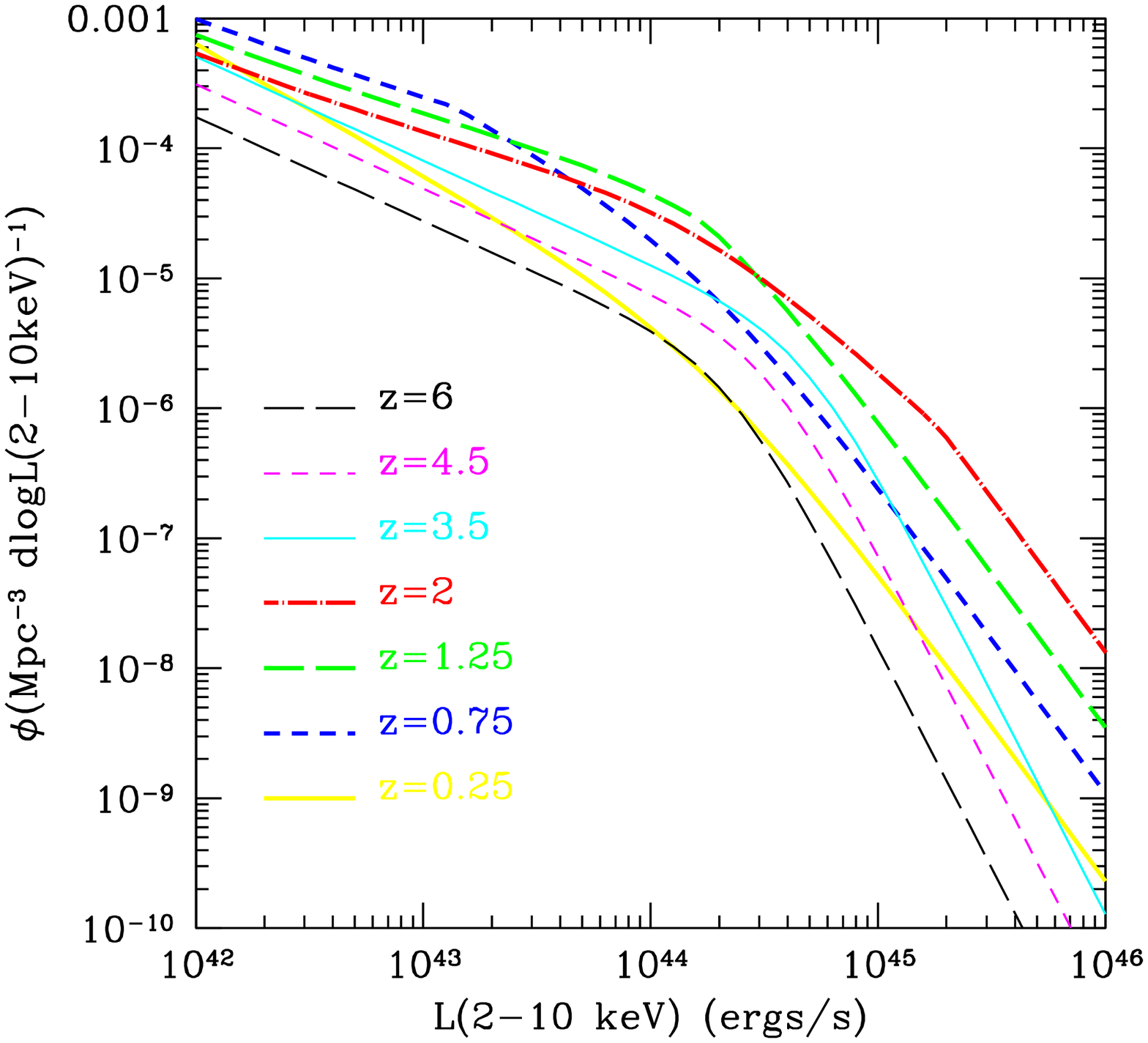}
\includegraphics[width=7cm]{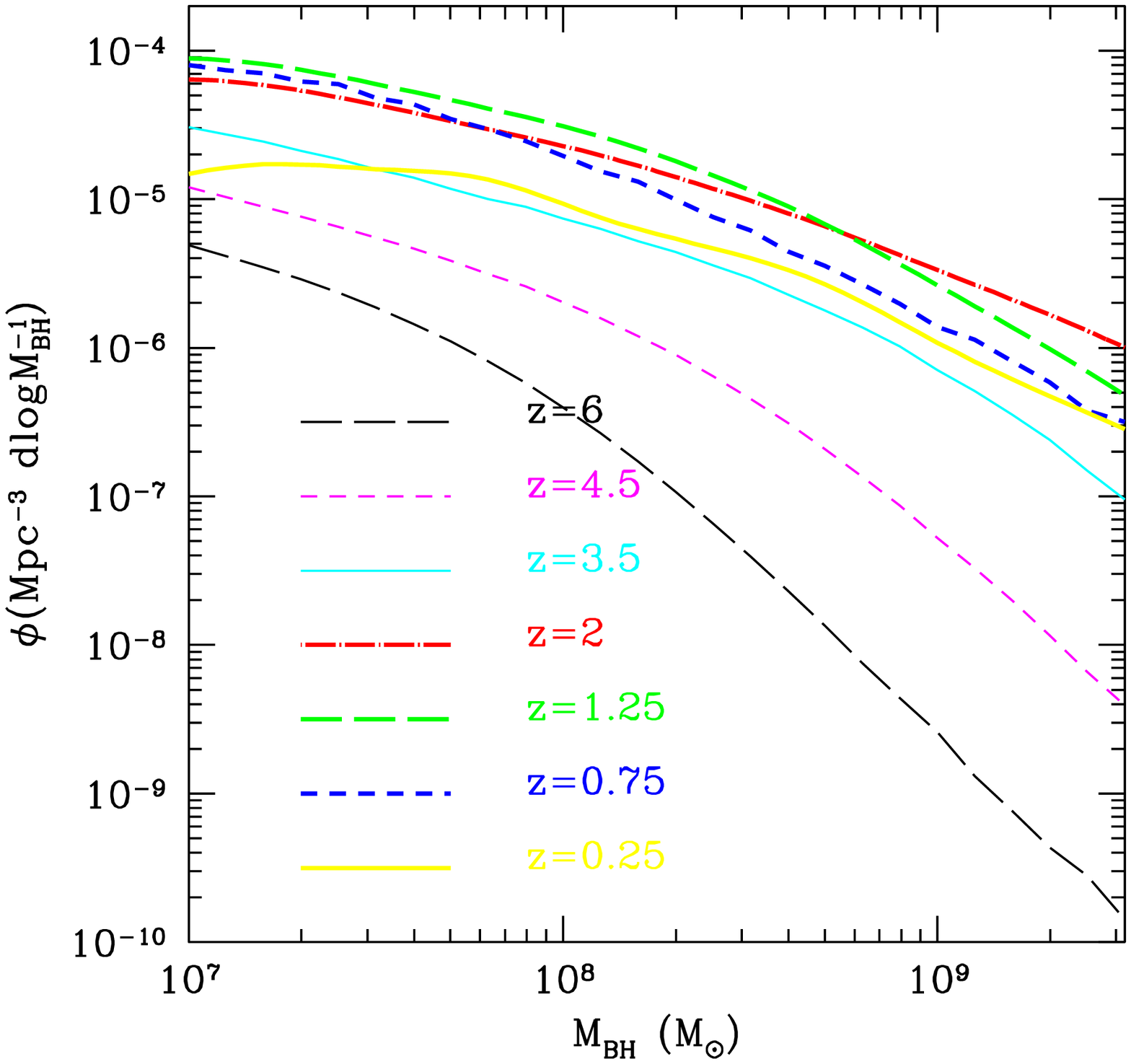}
\includegraphics[width=7cm]{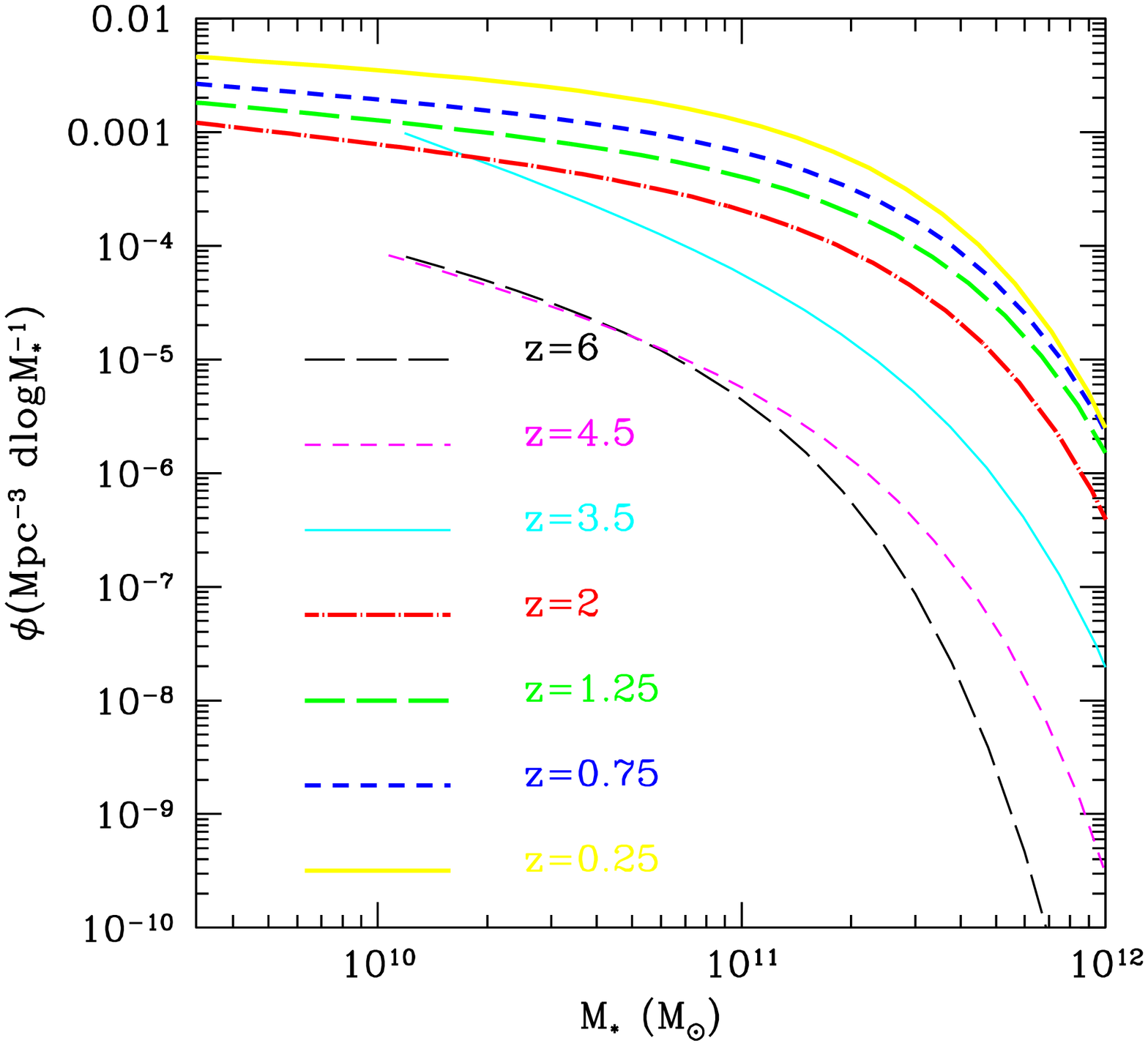}
\caption{[Top panel:] A collection of AGN luminosity
  functions. z$\ls2$ from La Franca et al. 2005, z$>3$ this
  work. [Central panel:] SMBH mass functions obtained by combining
  the AGN luminosity functions in the top panel with Eddington ratio
  distributions (see the text for details).  [Bottom panel:] A
  collection of galaxy stellar mass functions. z$\ls2$ from Fontana et
  al. 2006, z=3-4 from Santini et al. 2011, z=4-5 from Caputi et
  al. 2011, z=6 from Stark et al. 2009.}
\label{lfmf}%
\end{figure}

It is interesting to put the findings of the previous section in a
context, to study the evolution of the AGN luminosity function from
the local Universe to z$\sim6$.  We compare in the upper panel of Fig.
\ref{lfmf} the best fits to the z=3.5, z=4.5 and z=6 AGN luminosity
functions (model 4 in Table 5) to the best fit model luminosity
functions found at lower redshift by \cite{lafranca:2005}.  These have
been computed by correcting for the incompleteness due to X-ray
absorption, which is important at low redshift where the photoelectric
cut-off produced by the typical column densities observed in many AGN
is well within the X-ray selection band. The \cite{lafranca:2005}
luminosity functions are consistent with the new determinations of
\cite{aird:2010} and \cite{ebrero:2009}, based on a larger number of
objects.  The upper panel of Fig. \ref{lfmf} summarizes the complex
evolution of the AGN luminosity function, which rises from z=0 up to
z=1, 2, 3 for increasing luminosities and than decreases at higher
redshifts.

We converted these luminosity functions into ``active'' SMBH mass
functions by using MonteCarlo realizations. We simulated $10^8$ AGN
luminosities and SMBH masses in each redshift bin according to the
following procedure. 1) We first randomly choose an X-ray luminosity
following the luminosity function distribution in each given redshift
bin; 2) we than convert it into a bolometric luminosity using the
\cite{marconi:2004} and Sirigu et al. 2011 luminosity dependent
bolometric correction; 3) we randomly choose an Eddington ratio from
given distributions.  We used log-normal distributions with parameters
in Table 6. At z$<0.3$ we used the distribution of
\cite{netzer:2009b}, which are shifted toward higher Eddington ratios
with respect to the \cite{kauffmann:2009} distributions (see the
discussion in \cite{netzer:2009b,netzer:2009a}).  At medium to high
redshift we used the distributions of
\cite{trakhtenbrot:2011,shemmer:2004,netzer:2007,willott:2010b}. This
provides a SMBH mass for each chosen X-ray luminosity and Eddington
ratio. 4) We finally binned the resulting SMBH distributions in each
redshift bin to build SMBH mass functions. The resulting ``active''
SMBH mass functions in six redshift bins are plotted in the central
panel of Fig \ref{lfmf}. It must be stressed that this is an empirical
calculation, performed independently in each redshift bins, using only
observed quantities. It does not pretend to model the history of SMBH
growth through the cosmic times, which can be obtained by using a
continuity equation, conserving the number, as in
\cite{marconi:2004,merloni:2008,shankar:2009a}. 

Furthermore, the adopted Eddington ratio distributions all refer to
relatively luminous AGN.  Less active SMBH, giving rise to low
luminosity AGN and LINERs are not represented in these distributions.
For this reason we adopted a luminosity limit for the computed SMBH
mass functions: these functions represent ``active'' SMBH producing an
X-ray luminosity $>10^{43}$ ergs/s.  As a consequence, comparison with
previous calculation is not straightforward.  For example, while our
SMBH mass functions are luminosity limited, the ``active'' SMBH mass
functions of \cite{merloni:2008} are X-ray flux limited mass
functions. The 2-10 luminosity corresponding to the faintest flux
limit of \cite{merloni:2008} is $\sim10^{40}$ ergs/s at z=0.1,
$\sim10^{43.1}$ and $\sim10^{43.5}$ at z=3, while we plot the SMBH
mass function of AGN more luminous than $\sim10^{43}$ in all redshift
bins. SMBH mass functions of broad line AGN have been computed by
Kelly et al. 2010, by using about 10.000 SDSS QSOs in the redshift
range 1--4.5. Our SMBH mass functions are consistent, or slightly
higher than the Kelly et al. determinations at most redshifts and
black hole masses considered, as expected since broad lines AGN are a
fraction of the full active SMBH population, even at the highest SMBH
masses and/or AGN luminosities. The only regime where our estimates
are significantly higher than the Kelly et al. ones are the highest
masses (logM$>9.3$M$_{\odot}$) at z$\sim1$, where the Kelly et
al. 2010 function drops steeply, while our function decreases more
smoothly.

The SMBH mass functions can be transformed into stellar mass function
of ``active'' galaxies by assuming a conversion factor between SMBH
mass and host galaxy stellar mass.  We assumed a mean value
$\Gamma_0=log(M_{BH}/M_*)=-2.8$ at z$\sim0$ (\cite{haring:2004} but
also see the discussion in \cite{lamastra:2010}) and a redshift
evolution $\Gamma\sim \Gamma_0\times (1+z)^{0.5}$
(\cite{merloni:2004,hopkins:2006,shankar:09}). It should be noted that
in the local Universe the above correlation has been found and
calibrated for the bulge component of galaxies. Whether a strict
distinction between bulge and disk still exist at high-z is a matter of
debate. Disks of z$\sim2$ galaxies are much more compact and much
thicker that in today spirals of similar mass
(\cite{genzel:2006,vanderwel:2011}). For the sake of simplicity we
assume in the following calculation that the SMBH mass is proportional
to half of the total stellar mass of high-z galaxies. It is
instructive to compare the SMBH mass functions and the ``active''
galaxies stellar mass functions to the stellar mass functions of all
galaxies.  The lower panel of Fig. \ref{lfmf} plots a collections of
galaxy stellar mass functions: we used the stellar mass functions of
\cite{fontana:2006} at z$\ls2$, \cite{santini:2011} at z=3-4,
\cite{caputi:2011} at z=4-5 and \cite{stark:2009} at z=6.

\begin{table}
\caption{\bf Parameters of log-normal Eddington ratio distributions}
\begin{tabular}{lcc}
\hline
z  & peak    & width \\
   &         &  dex \\
\hline
0-0.3    & 0.03 & 0.5   \\
0.3-1    & 0.1  & 0.33 \\
1.5-2.5  & 0.18 & 0.33 \\
3-4      & 0.22 & 0.33 \\
4-5      & 0.5  & 0.33 \\
$>5.8$   & 1.0 & 0.25 \\
\hline
\end{tabular}

\end{table}

The AGN fraction, or AGN duty cycle, can finally be obtained by
dividing the ``active'' galaxy stellar mass functions by the galaxy
stellar mass functions.  Following the adopted luminosity limit used
to compute SMBH mass functions, we define AGN duty cycle the fraction
of AGN with 2-10 keV luminosity greater than $10^{43}$ ergs/s to the
total number of galaxies with a given stellar mass. Lower luminosity
AGN do exist and are probed by the luminosity functions in Fig.
\ref{lfmf} up to z$\sim2-3$. However, they are below the flux limit of
current surveys at higher redshift. Therefore, our luminosity
threshold also avoids problems of incompleteness of the samples at
high-z. According to this definition, the AGN duty cycle can be
  different from the AGN timescale, since the latter is correlated
  with the total inntrinsic lifetime of the AGN, which can include
  phases with luminosity $<10^{43}$ ergs/s.  The AGN duty cycle is
plotted in the upper panel of Fig. \ref{agndc}. It must be stressed
again that this is an empirical calculation, performed independently
in each redshift bin, using only the observed AGN and galaxy
luminosity and mass functions and observed Eddington ratio
distributions. We find that the AGN duty cycle increases at all
redshift with the stellar mass. Similar trends have been found by
\cite{kauffmann:2003,best:2005,bundy:2008,yamada:2009,brusa:2009b} for
optically selected, radio selected and X-ray selected AGN.  At each
given stellar mass the duty cycle increases with the redshift up to
z=4-5, consistent with the result of
\cite{marconi:2004,brusa:2009b,shankar:2009a}.  There are 22 galaxies
in the GOODS-ERS catalog with z$>3$ and stellar mass higher than
$10^{11.25}$ M$_\odot$, 7 of which are X-ray sources in Table 1, thus
confirming an AGN duty cycle $\gs30\%$ at z$>$3 for such massive
galaxies.  Fig. \ref{agndc}, lower panel, plots the evolution of the
AGN duty cycle as a function of the redshift for two galaxy stellar
masses: log($M_s)$=11.25 and 11.75. Bands are plotted rather than
curves to emphasize the rather large uncertainties, especially at
high-z (see next section).  The same Fig.  shows previous evaluations
by \cite{brusa:2009b}. These are consistent with the present estimates
within their rather large error bars.  The expectations of the
\cite{menci:2008} semi-analytic model are also showed in the lower
panel of Fig. \ref{agndc}.

\subsubsection{Error budgets}

The determination of the evolution of the AGN duty cycle is plagued by
large uncertainties, especially at high redshifts. It is therefore
important to study in detail the origin of these uncertainties and the
way of reducing them.

The uncertainty on both the AGN comoving densities at z$<$2 (at least
for unobscured and moderately obscured AGN) and on the stellar mass
functions at z$<2$ are relatively small, of the order of 10-20\% or
even smaller, thanks to large AGN and galaxy samples used for these
determinations and thanks to the use of several different surveys,
which helps in reducing the systematic error. The largest uncertainty
in the AGN luminosity function is on the fraction of Compton thick
AGN.  The fraction of these objects in the local Universe is high:
30-50\% of the optically selected Seyfert 2 galaxies can be Compton
thick ($\approx 1/4-1/3$ of the full AGN population,
\cite{risaliti:1999,panessa:2006}, a result confirmed by hard X-ray
selection, see \cite{malizia:2009}).  At higher redshift the fraction
of Compton thick AGN is more uncertain but it can be as high as 1/3 of
the full AGN population (\cite{fiore:2008,fiore:2009}), or even higher
(\cite{daddi:2007,treister:2010}).  A fraction of Compton thick
sources $\sim 1/3$ of the total is actually included in the model of
\cite{lafranca:2005}, and therefore we are confident that the error on
the total AGN comoving space density at z$<2$ due to undetected, and
unaccounted for, Compton thick sources is small ($\ls10-20\%$)
 
At z$>3$ the uncertainties on both AGN luminosity function and galaxy
mass functions are larger, in particular at low AGN luminosities and
low galaxy masses (see Table 3), they are of the order of 30-50\% at
z=3-5. At z$\sim6$ the uncertainties on the faint end of the AGN
luminosity function and on the galaxy stellar mass function are
extremely large, a factor 100\%.  These uncertainties are dominated by
statistics in the case of AGN (only few detections at z$>$4-5) and by
systematics in the case of the galaxy mass function (at z$\sim6$
stellar masses are estimated by \cite{stark:2009} by converting the UV
luminosity into stellar mass, which is a highly uncertain procedure).
The uncertainty due to undetected or unaccounted for Compton thick
AGN at z$>3$ is smaller than the above features.  In fact, we note
that at z$>3$ the cut-off produced by a column density $10^{24}$
cm$^{-2}$ is shifted below 2.5 keV, where the effective area of
Chandra and XMM peaks. This, together with the extremely deep
exposures of the CDFS, helps in detecting directly at least mildly
Compton thick sources
(\cite{georgantopoulos:2009,georgantopoulos:2011,feruglio:2011,gilli:2011,comastri:2011}).
Indeed, about one fifth of the Chandra-GOODS-ERS sources can be
Compton thick, as well as one fourth of the Chandra-GOODS-MUSIC
sources with optical spectroscopy (see Sect. 3.2), and many are
directly detected in our z$>$3 search.  We are therefore confident
that the uncertainty on the total AGN comoving density linked to
Compton thick AGN is at most $\sim10-20\%$ also at z$>3$. The high
luminosity end of our z$>$3 luminosity functions is probed by optical
surveys. Obscured AGN may well be missed in these surveys, and
therefore these determinations may be regarded as lower limits. We
note, however, that the fraction of obscured AGN decreases strongly
with the luminosity (at least at z$<2-3$ where large area X-ray
surveys provided samples of high luminosity QSOs), and therefore the
error on the comoving space density of high luminosity AGN at z$>3$ is
also likely to be small ($<20\%$).  We should consider the uncertainty
in the conversion from UV to X-ray luminosities for the optically
selected AGN luminosity functions. This uncertainty is probably of the
order of 30\% (see the discussion in Shankar et al. 2009a).

Another relatively large source of error in the evaluation of the AGN
duty cycle is represented by the uncertainty on the Eddington ratio
distributions. We performed several tests using parameters slightly
different than those in Table 4. For these AGN we found that the duty
cycle (at log$M_*$=11) changes by a factor of 15-30\% and $\sim100$\%
for distribution peak and width differing by 30\% from the assumed
ones. The biggest effect is introduced by the width of the Eddington
ratio distributions: the broader the distribution the higher is the
resulting AGN duty cycle. The uncertainty on the Eddington ratios is
particularly severe for low luminosity, logL(2-10keV)$<43$ AGN. Since
this is also, roughly, the luminosity limit of our z$>3$ sample, we
decided to limit the analysis to AGN with logL(2-10keV)$>43$ ergs/s
only, and assume conservatively a 50\% (total) relative error on the
AGN duty cycle arising from the uncertainty on Eddington ratios.

\begin{figure}
\centering
\includegraphics[width=8cm]{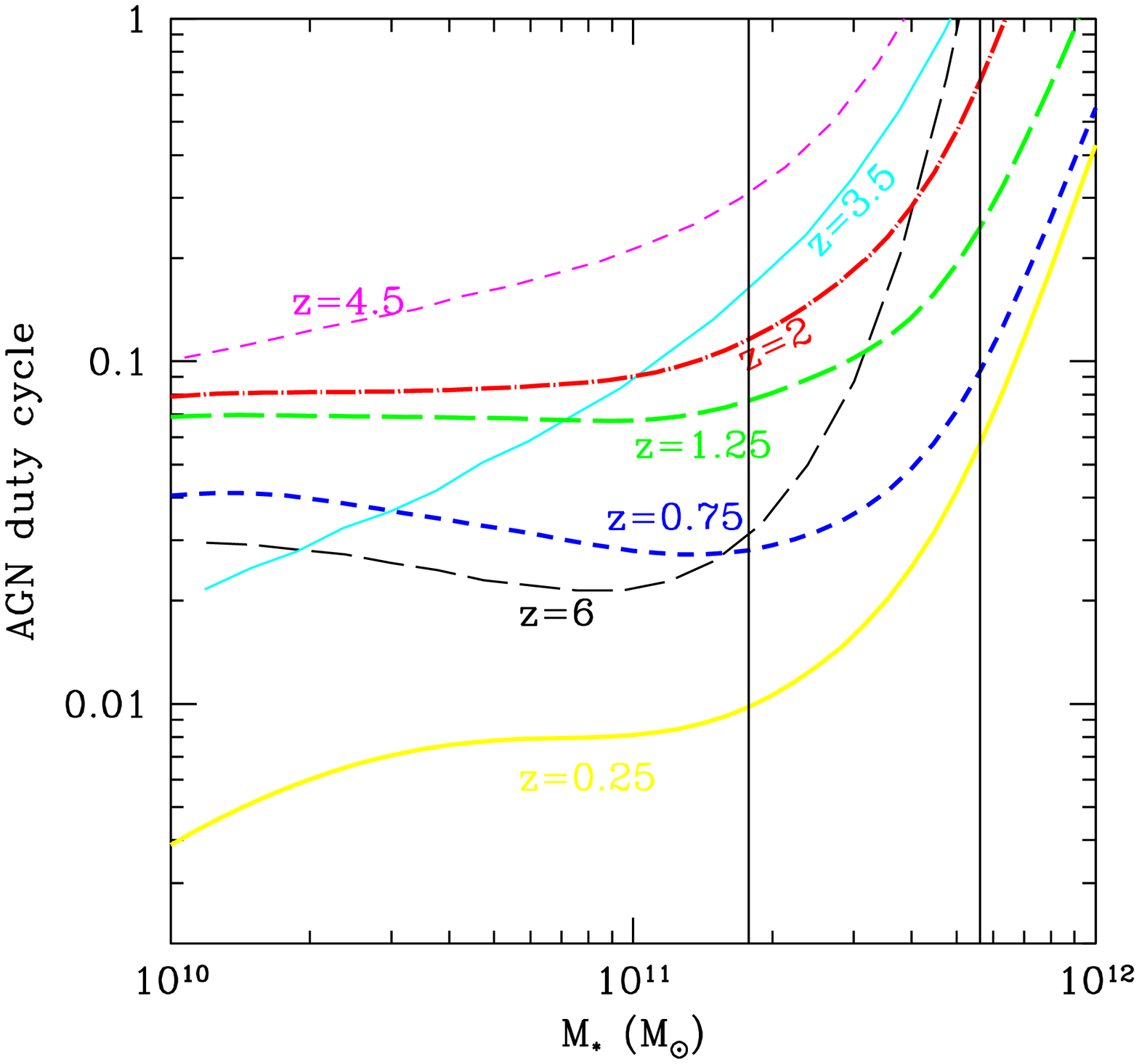}
\includegraphics[width=8cm]{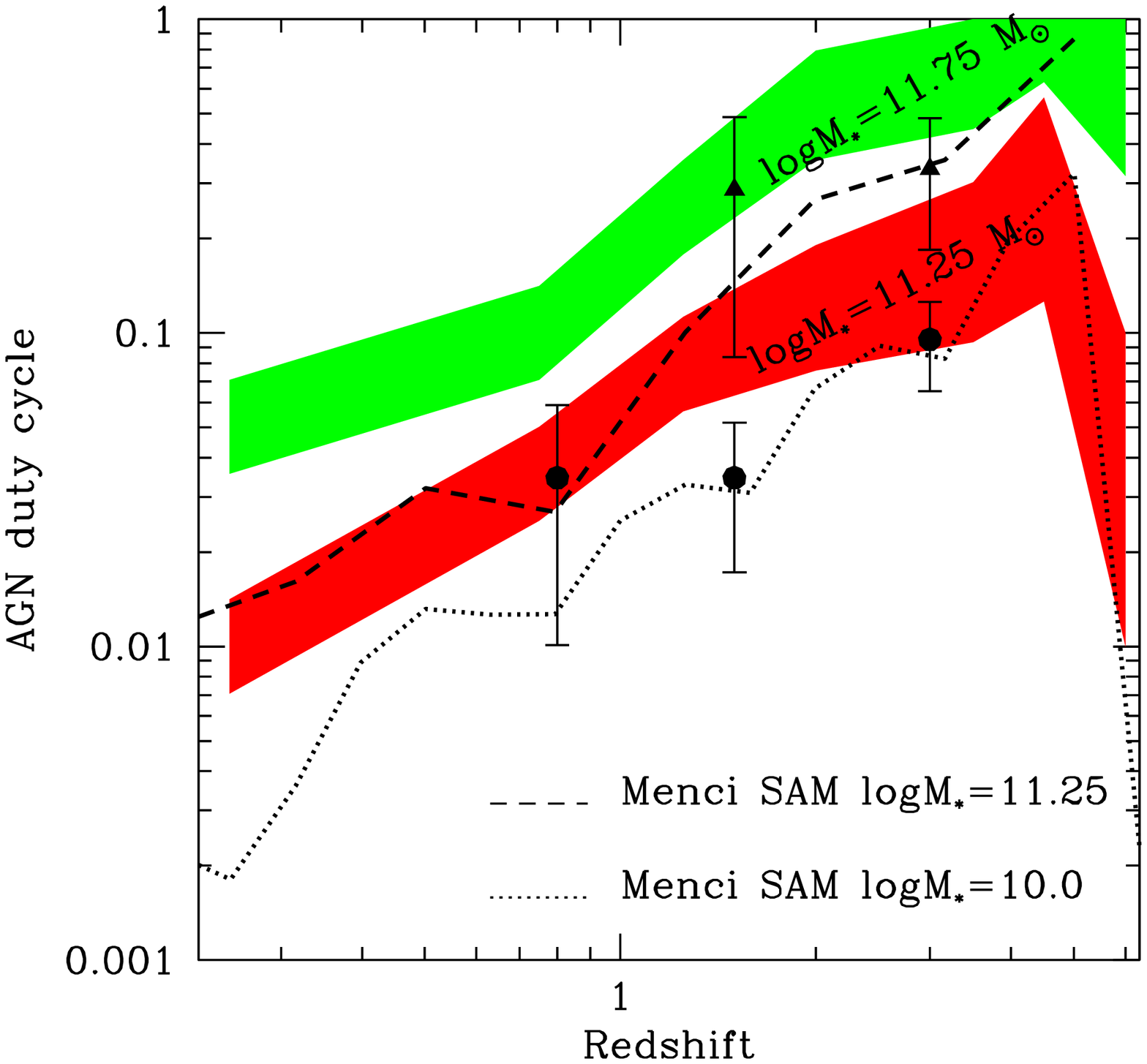}
\caption{[Top panel:] the AGN duty cycle as a function of the galaxy
  stellar mass in six redshift bins. AGN with logL(2-10keV)$>43$ are
  considered only. [Bottom panel:] the AGN duty cycle as a function of
  the redshift for two galaxy stellar masses (logM$_*$=11.25, 11.75
  M$_{\odot}$ ).  Filled circles and triangles are previous
  determination by Brusa et al. 2009 for the same masses. The black
  dashed (dotted) curve is the expectation of the Menci et
  al.(2006,2008) SAM for logM$_*$=11.25 M$_{\odot}$ (logM$_*$=10
  M$_{\odot}$ ). }
\label{agndc}%
\end{figure}

Finally, we adopted given normalization and evolution of the SMBH to
galaxy stellar mass ratio $\Gamma$ and assumed that is is the same for
each object at each given redshift.  However, we know that, at least
at low redshift, the SMBH-bulge mass relation is unlikely to be
universal (\cite{mathur:2001, mathur:2011} and references
therein). Furthermore, \cite{batcheldor:2010} suggest that selection
effects are important in shaping the SMBH-bulge mass relation.  Thus,
the deviations in the SMBH mass-bulge mass relations resulting from
galaxy morphology, orientation and selection effects can affect the
duty cycle calculations, in particular at small masses. We assumed a
redshift evolution $(1+z)^{0.5}$, consistent with our present
knowledge and with the expectation of several models
\cite{lamastra:2010,hopkins:2006,shankar:2009a}.  However, the
uncertainty on this calibration increases with the redshift, and can
be rather large at z$>3-4$.  To quantify all these effects is not an
easy task. As an example, if $\Gamma$ is half of what assumed above,
the duty cycle is reduced by a factor between 20\% and 100\%,
depending on the galaxy stellar mass and redshift.  In summary, the
bands plotted in Fig. \ref{agndc} roughly account for the typical
errors given above at each given redshifts and galaxy stellar masses.

\section{Discussion}

We evaluated the comoving space density of the z$>3$ faint X-ray
sources in three redshift bins: 3-4, 4-5 and $>5.8$. The number of AGN
in the three redshift bins is small, 19, 9 and 2 respectively, and
therefore we were forced to use relatively wide luminosity bins, to
keep the statistical error reasonably small. In particular, the
comoving space density at z$>5.8$ and logL(2-10keV)=43.5-44.5 has been
computed using only two sources. We stress that in one case a
secondary solution in the photometric redshift does exist at z$\sim2$,
so our determination is probably an upper limit to the true space
density of low luminosity AGN at $>5.8$.  This confirms that the slope
of the faint end of the z$>5.8$ AGN luminosity function is
significantly flatter than the bright end slope (see
e.g. \cite{shankar:2007}).  We also note that the other claimed z$>7$
AGN in the CDFS (L306, M70437), has a 0.5-2 keV flux below our
threshold for Chandra-GOODS-MUSIC sources, and that the photometric
solution of GOODS-MUSIC is broader than \cite{luo:2010} one, with a
lower limit at z=2.7.  This source is therefore {\it not} part of the
$>5.8$ sample used in Fig. \ref{hzlf}. Recently
  \cite{treister:2011} evaluated the integrated AGN emissivity at
  z$\sim6$ by stacking together the CDFS and CDFN X-ray data at the
  position of the z$\sim6$ \cite{bouwens:2006} galaxy candidates. They
  obtained a luminosity density of $1.6\times10^{46}$ ergs/s/deg$^2$
  in the 2-10 keV band. If we integrate our z$\sim6$ best fit model
  above logL$_X$=42 we obtain a total luminosity density of
  $7.5\times10^{38}$ ergs/s/Mpc$^3$ or $5.6\times10^{45}$
  ergs/s/deg$^2$, a value $\sim2.8$ times smaller than that reported
  by \cite{treister:2011}. It must be noted that our determination is
  based on a fit to the broad luminosity range luminosity function
  obtained by joining single X-ray detections in the CDFS to the UV
  luminosity functions, while the \cite{treister:2011} is based on a
  stacking analysis of faint X-ray sources in both CDFS and CDFN.

Our results put some first constraints on the faint end of the AGN
luminosity function (42.75$<$logL(2-10keV)$<$44.5) at z$=3-7$, which
are already interesting to be compared to model predictions.  To constrain
the shape of the AGN luminosity function we joined our determinations
with those obtained at higher luminosities by shallower X-ray surveys
(Chandra-COSMOS, XMM-COSMOS), and optical surveys (SDSS, NOAO
DWFS/DLS). We then compared the broad luminosity range luminosity
functions with the prediction of the semi-analytic model (SAM) of
\cite{menci:2006,menci:2008}.  In this model baryonic processes are
associated with the merging histories of dark matter (DM)
haloes. These haloes contain hot gas at virial temperature, a fraction
of which can radiatively cool down and form a disk with radius $r_d$
and circular velocity $v_d$. During mergers, and also looser galaxy
encounters (\cite{cavaliere:2000}), cold gas can loose its angular
momentum and can be accreted by the nucleus. A fraction of this gas
fuels a nuclear starburst while the rest can accrete on a central
SMBH, giving rise to an AGN. The AGN timescale is $\tau\approx
r_d/v_d$, i.e.  the crossing time for the destabilized gas. Assuming
typical values $r_d\approx$a few kpc and $v_d\approx 100$ km/s the AGN
timescale is short, a few $10^7$ yr, comparable with the Salpeter
timescale. In this SAM the AGN timescale, as well as AGN SMBH masses
and Eddington ratios, are {\it not} free parameters, but are
calculated self-consistently (the model only assumes that the
accretion can proceed at most at the Eddington limit). The SAM
includes a rather detailed treatment of AGN `quasar mode' feedback
(\cite{menci:2008}), in terms of a blast wave carrying outwards the AGN
power (\cite{lapi:2005}). The SAM produces prediction for all AGN,
both unobscured, moderately obscured and Compton thick.
\cite{lamastra:2010} compares the SMBH masses and stellar masses of
AGN host galaxies predicted by this SAM with measurements for various
galaxy and AGN samples at different redshifts.  This SAM reproduced
reasonably well the z=3-4 luminosity function at
logL(2-10keV)$>43.5$. At lower luminosities it predicts 2-3 times more
AGN than observed. It should however be noted that some extreme
Compton thick AGN, those in which the nuclear emission is completely
blocked by photoelectric absorption and Compton scattering, leaving
only reflection emission in the X-ray band, can well be missed even in
the deepest X-ray surveys. The agreement is sufficiently good at high
luminosities (logL(2-10keV)$>45$) at z=4-5 and z$>5.8$ too. In these
redshift bins the agreement between data and model gets worse at lower
luminosities, where the \cite{menci:2008} SAM predicts 5-10 times more
low luminosity AGN than in our determination and 2-3 times more than the
\cite{treister:2011} ones.

To investigate the origin of this behaviour and get insights on high-z
AGN physics we also compared our measurements with predictions derived
from more basic models for AGN activation through galaxy interactions.
This class of models consists of two ingredients: a DM halo merger
rate compatible with cosmological simulations, and an input AGN light
curve (e.g., Wyithe \& Loeb 2003, Lapi et al. 2006, Shen 2009, Shankar
2009, 2010, and references therein).  The initial mass of the SMBH at
triggering is assumed to be a fixed small fraction of its mass at the
peak of activity. SMBH growth is regulated by a condition between the
peak luminosity and the mass of the host halo at triggering,
consistent with the local relations between SMBHs and their host
galaxies.  The main advantage of approaching AGN modeling through
this simplified technique is that not being part of a specific SAM,
they can explore easily and quickly a large space of parameters and
physical recipes to trigger AGN. Here we follow the models presented
in the preliminary work by Shankar (2010a) with same parameters as in
Shen (2009). A more comprehensive and detailed analysis of AGN merger
models, with implications for SMBH scaling relations and AGN
clustering (Shankar et al. 2010a,b) is beyond the scope of the present
paper and will be discussed elsewhere (Shankar et al. 2011, in
preparation).  In this model AGN are activated by mergers of the host
DM haloes ($\xi=$M$_{h2}/$M$_{h1}>0.3$ for major mergers, $\xi>0.1$ includes
minor mergers too). The AGN is triggered with no dynamical friction
time delay between host halo and actual galaxy-galaxy merging (a
solution close to the fly-by hypothesis of
\cite{cavaliere:2000}). Super-Eddington accretion ($L/L_{Edd}=3$) is
allowed in the initial phases of BH growth, and a long, sub-Eddington
post-peak phase is present in each event. Finally, a minimum halo mass
M$_{hmin}$ is assumed (see Shen 2009 for details).  We plot in the
central and lower panel of Fig. \ref{hzlf} four different models:

\begin{enumerate}
\item
$\xi>0.3$, $L/L_{\rm Edd}=3$,  M$_{hmin}=3\times 10^{11}\,  M_{\odot}/h$ (blue, solid lines, reference model);  
\item
$\xi>0.1$, $L/L_{\rm Edd}=1$,  M$_{hmin}=3\times 10^{11}\,  M_{\odot}/h$ (blue, dotted lines);  
\item
$\xi>0.3$, $L/L_{\rm Edd}=3$,  M$_{hmin}=3\times 10^{11}\,  M_{\odot}/h$, without descending phase (blue, thick dashed lines);  
\item
$\xi>0.3$, $L/L_{\rm Edd}=1$,  M$_{hmin}=0.1\times 10^{11}\,  M_{\odot}/h$ (blue, long-dashed lines);  
\end{enumerate}

The purpose of the comparison of these models with the data is to
understand whether the merger scenario produces expectations in
agreement with the data and to highlight which of the underlying
physical assumptions makes this match actually possible.

The reference model (1) overpredicts by a factor of 2-3 the AGN
luminosity functions at z$>$4 at all luminosities but
logL(2-10keV)=43.5-44.5 at z$>5.8$. The same model without a
descending phase (3) reproduces relatively well both the z=4-5 and the
z$>5.8$ luminosity functions. This suggests that the underlying AGN
light curve must not be too prolonged in order not to generate too
numerous faint, sub-Eddington AGN.

The inclusion of minor mergers steepens the AGN luminosity function at
both z=4-5 and z$>5.8$. Model (2) with $\xi>0.1$ and $L/L_{Edd}=1$ is
roughly consistent with the z=4-5 luminosity function but
underpredicts the bright end of the z$>5.8$ luminosity
function.

Relaxing the assumption of a minimum halo mass steepens again the AGN
luminosity function at both z=4-5 and z$>5.8$. The same model with
$L/L_{Edd}=1$ (model 4) overpredicts the low luminosity end of the z=4-5
luminosity function and underpredicts the high luminosity end of the
z$>5.8$ luminosity functions.

In conclusion, we find that the inclusion of minor mergers and the
exclusion of a minimum halo mass both steepen the model luminosity
function.  About the first, minor mergers and galaxy fly-by helps the
formation of stars at early times in the progenitors of today massive
galaxies, thus helping in explaining their observed colors
(e.g. \cite{menci:2004}). They also help in reproducing quite well the
X-ray AGN space densities at z$\ls3$ (\cite{menci:2008}).  About the
latter, there could be several causes for a minimum halo mass and/or
for inefficiency of BH formation and growth in low mass halos. For
example, in the SMBH formation scenario where seed BHs are formed from
the monolithic collaps of gas clouds, the probability for a DM halo of
hosting a BH seed is an increasing function of the halo mass
(\cite{volonteri:2008}). On the other hand, in the SMBH scenario where
BH seeds are formed from PopIII stars, basically all DM halos with
M$_h>10^{9-10}$ M$_\odot$ at z$<7$ are populated with BHs. Therefore,
at least in principle, the faint end of the high-z AGN luminosity
function might be used to discriminate among these two
scenarios. However, there are at least other three effects that
produce a minimum halo mass for BH formation and growth.  First, the
gravitational recoil, giving rise to ejection of one of the BH after a
merging, can be more important at low masses (see
e.g. \cite{volonteri:2011}). Second, a cut-off in the build up of
galaxies in haloes with M$_h<10^{11}$ M$_\odot$ could be produced by a
metallicity effect (\cite{krumholz:2011}).  Third, an intrinsic
cut-off at low halo masses can be due to a cut-off in the primordial
power spectrum of gravitational perturbation at small scales. Free
streaming of {\it warm} DM leeds to a suppression on scales smaller
than the free streaming scale and therefore to a cut-off. {\it Warm}
DM has often been invoked to solve problems of the standard $\Lambda
CDM$ scenarios at small scales, such as the large number of galactic
satellites, the cuspiness of galactic cores, and the large number of
small, compact galaxies predicted by these models (see
e.g. \cite{primack:2009}). However, the limits on the cut-off scale
(or, equivalently, on the mass of the {\it warm} DM particle) derived
from high-z Ly$\alpha$ forest are quite stringent: up to $\sim100$ kpc
no cut-off is seen (\cite{viel:2008} and references therein). It is
difficult to asses quantitatively which are the DM halo scales
influenced by these effect. To this purpose we should include in our
SAM and model both gravitational recoil and {\it warm} DM, and compare
their prediction with AGN and galaxy luminosity functions. This is
beyond the scope of this work and will be part of a separate
publication (Menci et al. 2011, in preparation). A minimum halo mass
for black hole growth implies that accretion is inhibited in halos
smaller than this threshold mass ($3\times10^{11}$ M$_{\odot}$ in our
models). BH seeds in these halos do not grow, until they are merged to
BH in bigger halos. As a consequence, nuclear accretion occurs only on
a fraction of the merging history, and the halo population when the
above threshold is reached late will host SMBH with small
masses. Qualitatively, they will be somewhat below the {\em Magorrian}
correlation. Again, to obtain quantitative estimates we need to study
the predictions of models including physical causes for the cut-off in
the halo mass, which will be addressed in a forthcoming pubblication.

Finally, we have joined our present determination of the high-z AGN
luminosity function with previous determination at lower redshift and
with galaxy mass function determinations, to empirically estimate the
AGN fraction (or AGN duty cycle), as a function of the galaxy stellar
mass and redshift.  We find that the AGN duty cycle increases with
both host galaxy stellar mass and redshift. The AGN duty cycle is
computed by assuming a given evolution of the {\em Magorrian}
  relationship.  and by assuming distributions of Eddington ratios at
different redshifts (see Table 6).  These are far to be accurately
known. Indeed the distributions of Table 6 taken from
\cite{trakhtenbrot:2011,shemmer:2004,netzer:2007,willott:2010b} show an
evolution with the redshift, while this is not the case for the
distributions of \cite{kollmeier:2006}. However, the increase of AGN
duty cycle with redshift depends marginally on the evolution of the
Eddington ratio, reflecting more basic AGN population properties,
evidenced by the large evolution of the AGN number density (see e.g.
\cite{menci:2006,menci:2008,shankar:2009b} and references therein).
The correlation of AGN duty cycle with both stellar mass and
redshift is a rather robust result. It holds also by changing
the calibration adopted for the {\em Magorrian} relationship by a factor
of 2. Our calculation confirms what clustering measurements
suggest (e.g., \cite{shankar:2010a}): the large clustering
signal from luminous quasars implies that BHs reside in massive
haloes that by definition are rare and thus require high duty
cycles.

We compare in Fig. \ref{agndc} the evolution of the
AGN duty cycle with the prediction of the \cite{menci:2008}
SAM. We see that this model reproduce the observed trend of
increasing duty cycle from z$\sim 0$ to z$\sim5$, although the
observed AGN duty cycle at each redshift is somewhat smaller than
predicted for the same stellar mass. We recall again that Compton
thick AGN are present in the model, but a fraction may be missed
in the data. This rough agreement suggests that the triggering of
at least relatively luminous AGN (logL(2-10keV)$\gs$43 ergs/s),
is probably produced by galaxy interactions in the galaxy stellar
mass and redshift range investigated (up to z$\sim4-5$). At
higher redshifts the model predicts a saturation of the AGN duty
cycle of galaxies with masses logM$_*>11.25$ M$_\odot$ (i.e
basically all massive galaxies at z$\gs5$ are expected to host an
active nucleus with logL(2-10keV)$>43$ ergs/s).

\subsection{Future perspectives}

The faintest sources used in the comoving space density calculation
are from the Chandra-GOODS-ERS sample.  Indeed, we found that the
HST/WFC3 ERS H band images are deep enough to trace high-z AGN
populations at the extremely faint X-ray flux limits reached by the
Chandra 4 Msec exposure, avoiding significant incompleteness, or at
least with an incompleteness comparable to that reached at much
brighter fluxes by Chandra-COSMOS and XMM-COSMOS. This is not the case
for the HST/ACS optical and Spitzer IRAC images from which the
GOODS-MUSIC catalog is generated. These images are too shallow to find
the counterparts of all faint X-ray sources at the extremely deep CDFS
4 Msec flux limit. We conclude that to fully exploit ultradeep Chandra
data, i.e. to keep the identification rate of faint X-ray sources
comparable to that reached at brighter fluxes, a deep NIR (H$\gs26$)
coverage is mandatory. To avoid large incompleteness in the GOODS area
without HST/WFC3 coverage we were forced to use only X-ray sources
brighter than $5\times10^{-17}$ \cgs.  It must be noted that the ERS
area lies at relatively large off-axis angles in the Chandra field,
where the sensitivity of Chandra is reduced by a factor of at least 2
by both vignetting and point spread function (PSF) degradation. As a
result, we could not exploit in this work the full sensitivity of the
4 Msec Chandra field.  This limitation can be overcome in future, when
the HST/WFC3 CANDELS survey (Grogin et al. 2011; Koekemoer et al. 2011)
will be ultimated. The CANDELS survey covers the central part
of the Chandra field and therefore can allow us to fully exploit the
Chandra sensitivity. Based on the ERS results we expect to at least
double the sample of faint high-z AGN using the CANDELS survey of the
CDFS.

The CANDELS multy-cycle HST treasury program covers a total of five
fields: the CANDELS {\it deep} fields are in the GOODS-South and
GOODS-North regions, and cover a total of 130 arcmin$^2$ to a depth of
H=27.8. The CANDELS {\it wide} survey covers the AEGIS, COSMOS and UDS
fields for a total of 670 arcmin$^2$ to a depth of H$\sim26.5$
(similar or slightly shallower than in the ERS field). All these five
fields are going to be {\it premiere fields} for high-z galaxy and AGN
studies. If we extrapolate our finding on the ERS field to all CANDELS
fields, assuming a Chandra coverage of $\sim4$ Msec for the CANDELS
{\it deep} fields and of $\sim1$ Msec for the CANDELS {\it wide}
fields, we expect 30-100 z=4-5 AGN, 10-30 z=5-6 AGN and 3-10 z$>6$
AGN. This goal can be reasonably achieved by Chandra in the next few
years (i.e. adding $\sim5-6$ Msec of observation to $\sim8$ Msec
already spent on these field). Probably the biggest problem with this
sample would be the spectroscopic identification of these faint,
high-z AGN. A fraction can be spectroscopically confirmed by HST/WFC3
grism spectroscopy. For the faintest, the only viable chances are
probably ALMA observations in a few years, and, at the end of the
decade, JWST and ELTs observations.  At the end of this program we
should have a very good knowledge of the faint end of the AGN
luminosity function up to z=5, but a still rough determination of the
luminosity function at z=5-7. To make further progresses with Chandra,
i.e. quantitatively probe the first generation of accreting SMBH,
which would allow putting stringent constraints on SMBH formation
models
(\cite{madau:2001,lodato:2006,lodato:2007,volonteri:2010a,volonteri:2010b,begelman:2010}),
and accretion scenarios
(\cite{volonteri:2005,dotti:2010,fanidakis:2011,king:2008}), would
require to at least triple the exposure times, i.e. spend on deep
surveys other 30-40 Msec.  While this is not technically unfeasible,
it is certainly extremely expensive.  The Chandra limiting problem is
that its sensitivity is very good on axis, but degrades quickly at
off-axis angles higher than a few arcmin, making difficult and
expensive in terms of exposure time to cover with good sensitivity
area larger than a few hundreds arcmin$^2$.  A significant leap
forward in the field would then be obtained by an instrument capable
of reaching the Chandra Msec exposure, on axis sensitivity (i.e flux
limits in the range $1-3\times10^{-17}$ \cgs) but on a factor of $>10$
wider field of view. The extremely good Chandra PSF (HPD$\sim0.5$
arcsec), which is the main reason for the good Chandra sensitivity for
point sources, is beyond reach of today technology of thin X-ray
mirrors. However, this can be supplied for by increased throughput,
lower background and wider field of view with nearly constant PSF.  A
straw-man calculation shows that a telescope with the total XMM
throughput (i.e.  area $\sim$half square meter) and a PSF $\sim10$
times worse than Chandra (i.e. HPD$\sim5$ arcsec) but slowly degrading
with the off-axis angle, feeding a detector with half degree side and
2-3 time lower internal background would be as fast as Chandra but
with a $\sim20$ times better {\it Grasp}, i.e. sensitivity times field
of view. A further improvement could be achieved reaching an internal
background similar to that of Swift XRT on low Earth orbit ($\gs10$
times smaller than the Chandra internal background per unit solid
angle).

The situation at the bright end of the luminosity function is
fortunately easier. On this side, large areas will be surveyed in a
few years by the XMM XXL survey and by eROSITA. The XMM XXL survey
will cover $\sim25$ deg$^2$ of sky at a depth of $\sim 5 \times
10^{-15}$ \cgs with 10 ks effective exposure.  eROSITA will survey
most of the sky down to a 0.5-2 keV flux limit $\sim10^{-14}$ \cgs,
and hundreds deg$^2$ 2-3 times deeper. The identification of high-z
AGN in these surveys requires near infrared follow-up. This will be
possible with both present (VISTA) and future (Euclid)
instrumentation. In particular, Euclid will provide H=24 photometry on
most of the high galactic sky and H=26 photometry on about 50
deg$^2$. All these surveys will help in probing luminous AGN and test
the completeness of the optical large area surveys such as SDSS. They
will extremely useful to find rare high-z, high luminosity but highly
obscured QSOs.

The present data can only put rough constraints on the high-z AGN duty
cycle.  We stress that the AGN duty cycle determinations are presented
here for two main purposes: first we would like to point out that this
measure as a function of the redshift is, at least in principle, a
powerful tool to discriminate among different AGN triggering
mechanisms. The present data show a fast increase of AGN duty cycle in
massive galaxy with redshift, naively favoring galaxy encounters as
triggering mechanisms. However, the uncertainties are still too large
to draw any strong conclusion. Then, we use the AGN duty cycle
determination to emphasize which are the main areas of improvement to
reduce these uncertainties. First, we clearly need a better constrain
on the AGN logL(2-10keV)=43-44.5 density at z$>4$.  A major step
toward this direction can be obtained by using further combined
Chandra and HST surveys in the next few years.  Second, we need better
constraints on the galaxy stellar mass functions at z$>3$, in
particular the low mass end of the mass functions. Efforts in this
direction are on going, using both ground based surveys (with VISTA
and other near infrared cameras on 8m class telescopes) and from the
space (HST/WFC3 surveys and the Spitzer SEDS survey). Third, we need
better constraints on the Eddington ratio distributions as a function
of the redshift, in particular on the distribution at low
L/L$_{Edd}$. Last, we need to test additional physics, such as BH
gravitational recoil, and BH formation and growth in a {\it warm} DM
scenario, through the comparison of model predictions with observed
luminosity functions and AGN duty cycle.  On going and future Chandra
and HST surveys, in particular those planned in the CANDELS fields,
should be able to find statistically sound samples of z=$\gs5$ and
logL(2-10keV)$>43$ AGN (a few tens), and put reliable constraints on
the high-z galaxy stellar mass function, providing in turn much better
constraints on the AGN duty cycle.  Fig. \ref{agndc} also shows the
prediction of the \cite{menci:2006,menci:2008} SAM for low galaxy
stellar masses (M$_*\sim10^{10}$ M$_{\odot}$). For such masses the
duty cycle increase with the redshift, reach a maximum and then
decreases at z$\gs4$. This is due to the combination of two
effects. First, small galaxies form in relatively low density regions,
where gas cooling is slower, forming smaller gas reservoirs than for
large galaxies.  Second, at the highest redshifts the rate of galaxy
interaction is small, it increases up to the redshift were groups
forms and decrease again at low redshift because high relative
velocities and low densities.  The detection of this distinctive
expectation would be the unambiguous confirmation of the galaxy
interaction scenario for AGN triggering. While Chandra is sensitive
enough to probe $\sim10^{7}$ M$_\odot$ SMBH up to z$\sim7$, the low
mass end of the galaxy stellar mass functions at z$\gs4$ is today
practically unconstrained. The latter is therefore the area where
improvements are more needed, but unfortunately also more difficult
with the present generation of instruments.  A breakthrough in this
field would probably await for the next generation of infrared
telescope such as JWST and the ELTs.

\section{Summary}

We have exploited the ultra deep Chandra and HST coverage of the CDFS
to search for X-ray emission at the position of high redshift
galaxies. To this purpose, we have used {\it ephot}, a highly
optimized tool for X-ray source detection and photometry (see the
Appendix). We found significant emission at the position of 17
galaxies in the area covered by the HST/WFC3 ERS survey, and 41
galaxies in the full GOODS area for a total of 54 independent z$>3$
X-ray sources, 29 of which are not present in previous X-ray
catalogs. We reach a flux limit of $\sim1.7\times10^{-17}$ \cgs in the
0.5-2 keV band, which corresponds to luminosities of $\sim10^{42}$
ergs/s at z=3 and $\sim10^{43}$ ergs/s at z=7. The present Chandra
deepest exposure is thus able to probe Seyfert-like galaxies up to the
epoch of formation of the first galaxies and normal star-forming
galaxies up to z=3-4. Indeed we have evidence that the X-ray emission
of one of the faintest Chandra-GOODS-MUSIC source, M4417, is likely of
stellar origin. This is the source in our sample with the lowest X-ray
to H and z bands flux ratio, and the X-ray luminosity expected from
the SFR rate measured from the UV rest emission, oxygen lines and
radio 1.4 GHz emission is close to the observed one. This can be the
highest redshift inactive galaxy observed in X-rays.

We analyzed the X-ray spectra of the Chandra-GOODS-ERS sources finding
hard spectra in several cases. In 3 cases the spectra can be fitted
with both a reflection model or an absorption model with a column
density in excess $10^{24}$ cm$^{-2}$. In an additional case the
column density results in the range $10^{23}-10^{24}$. The fraction of
Compton thick AGN at 0.5-2 keV fluxes between 0.3 and
3$\times10^{-16}$\cgs and at z$>$3 is thus at least
$18^{+17}_{-10}\%$.  This is higher than predicted by popular models
for the CXB (\cite{gilli:2007}). Similar conclusions are reached by
\cite{gilli:2011} and \cite{treister:2011}. The optical counterparts
of the highly obscured and Compton thick AGN do not show any
reddening, and we thus conclude that the size of the X-ray absorber is
likely smaller than the dust sublimation radius ($\ls 1$pc, at the
luminosity of our AGN).

We compared the X-ray spectra of the Chandra-GOODS-ERS and
Chandra-GOODS-MUSIC sources with their optical spectra and optical/NIR
morphology. Two Compton thick or highly obscured AGN show broad
absorption lines in their UV rest frame spectra. Four of the ten
highly obscured and Compton thick AGN have a point like morphology in
their $z$ and H bands images, thus suggesting that the active nucleus
contributes significantly to the UV and optical rest-frame
emission. Indeed, their rest-frame UV (0.16$\mu$m) extinction,
estimated by fitting the observed SED with galaxy and AGN templates
turns out smaller than 1-2 magnitudes in all cases.  This implies that
the X-ray absorber must be practically dust free, a condition
satisfied if the X-ray absorbing matter is within or close to the dust
sublimation radius. This is a fraction of pc for our sources, a region
more compact than the pc-scale obscuring {\em torus}.  Compact
absorbers have been discovered in local Seyferts
(e.g. \cite{risaliti:2005}), they might be more common at high
redshift.

We searched for radio emission at the position of the Chandra-GOODS-ERS
and Chandra-GOODS-MUSIC sources in the deep VLA-CDFS maps at 1.4
GHz. We find highly significant emission (SNR$>$5) in 2 Chandra-GOODS-ERS
sources and 5 Chandra-GOODS-MUSIC sources.  We find radio emission
with lower signal to noise (2.4$<$SNR$<3.2$) in additional 8 sources
in Table 3. The presence of significant radio emission from the faint
X-ray sources is further confirmed by the analysis of the radio images
obtained by stacking together the radio images at the position of
Chandra-GOODS-ERS and Chandra-GOODS-MUSIC sources. The average flux per
source of 6-10 $\mu$Jy is higher than that expected by
\cite{lafranca:2010} based on the X-ray fluxes and assuming common
nuclear origin. This would suggest either a strong evolution of the
AGN radio ``loudness'' with the redshift (the \cite{lafranca:2010}
determination is mostly based on z$<3$ AGN), or that our high-z,
X-ray detected galaxies, are actively forming stars (SFR$>$a few
hundreds $M_{\odot}$/yr). In two cases the SFR evaluated from the
radio flux is consistent with that evaluated from sub-mm observations
(M208) or UV SED fitting (M4417).

We evaluated the comoving space density of the z$>3$ faint X-ray
sources in three redshift bins: 3-4, 4-5 and $>5.8$. We found that the
wide luminosity range z$>3$ luminosity functions are consistent with a
pure luminosity evolution model, with $L^*$ evolving quite rapidly,
and reducing by a factor 3.3 from z=3 to z=6: $L^*(z=3)=6.6\times
10^{44}$ ergs/s, $L^*(z=6)=2\times 10^{44}$ ergs/s. We compared these
space densities with the predictions of the
\cite{menci:2006,menci:2008} SAM and with the expectations of more
basic models for AGN activation through mergers.  We find that these
models are broadly able to reproduce the high-z AGN luminosity
functions. A better agreement is found by assuming a minimum halo
mass. We speculate that there are at least four effects or scenarios
that can produce a minimum halo mass for BH formation and growth: 1)
if BH are formed from the monolithic collaps of gas clouds, the
probability for a DM halo of hosting a BH seed is an increasing
function of the halo mass; 2) gravitational recoil, giving rise to
ejection of one of the BH after a merging, which can be more important
at low masses; 3) metallicity dependent star-formation; 4) a cut-off
in the primordial power spectrum of gravitational perturbation at
small scales. To assess the relative importance of all these effects
we need {\it both} better data and more complete models for galaxy and
AGN formation and evolution.

Finally, we empirically evaluated the evolution of the AGN duty cycle
(fraction of AGN more luminous than $10^{43}$ ergs/s in the 2-10 keV
band, to the total number of galaxies with the same mass) as a
function of both host galaxy stellar mass and redshift. We compared it
with the expectation of a model using galaxy interaction as AGN
triggering mechanism.  The uncertainties on the AGN duty cycle are
large at all redshifts. At z$<4$ they are dominated by the uncertainty
on the Eddington ratio distributions. At higher redshift the
uncertainty on both AGN luminosity function and stellar mass functions
are extremely large, in particular at low AGN luminosities and small
galaxy stellar masses. Nevertheless, it is clear how the AGN duty
cycle must change substantially from z=0-0.5 to z=4-5, increasing by
1.5 orders of magnitude for host galaxy stellar masses $1-5 \times
10^{11}$ M$_{\odot}$. This big rate of change of the AGN duty cycle of
massive galaxies is roughly consistent with that predicted by the
models using galaxy interactions as AGN triggering mechanism. We recall
again that this applies to relative luminous AGN (logL(2-10keV)$\gs$43
ergs/s). Triggering of lower luminosity AGN is of course not
constrained by our analysis.

\appendix
\section{{\it ePhot}: energy and space adaptive filter for source detection}

The standard techniques for source detections identify statistically
significant brightness enhancements, deriving from both unresolved
(point) and resolved (extended) X-ray sources, in images accumulated
in given energy ranges and observation times (such as tools
'celldetect', ´vptdetect' and 'wavdetect' in CIAO, 'emldetect' in
SAS). However, binning and projection always result in a loss of
information.  Conversely, present X-ray data, and in particular
Chandra/ACIS data, are incredibly rich, containing spatial, energy and
timing information at very good resolutions. Even though X-ray data
are organized in event files, preserving the information from the
datacube, the scientific analysis is usually not performed directly on
the original event files. The datacube is usually projected onto
spatial plane before performing source detection, and spectra are
accumulated at given times or at given sources positions before
performing spectral fits. To take full advantages of large datacubes
we should be able to detect sources in a multidimensional space. We
are developing multidimensional source detection techniques using
spatial, spectral and timing information.  We search for clustering of
X-ray events in a multidimensional space (energy and time in addition
to the usual spatial coordinates). This allows us to efficiently
detect X-ray bursty sources and X-ray sources characterized by a
peaked spectrum (strong lines, cut-offed spectra), by reducing the
elapsed time and/or the energy range (and therefore the background)
where the source detection is performed.

We start by accumulating spectra at each galaxy position using
different circular source extraction regions from 3 to 9 pixels on
cleaned event files.

As first step we selected all galaxies with more that 12 background
subtracted counts in the wide 0.3-4 keV band, to exclude the faintest
objects, and achieve a high reliability (see Sect. A.2 below). At this
stage the background is evaluated from a map obtained from the
original image after the exclusion of all bright sources (sources
detected by a wavelet algorithm at a conservative threshold).

For all these galaxies we search for the source extraction region
radius and contiguous energy band that minimize the Poisson
probability for background fluctuation and maximize the signal to
noise ratio. We adopt a logarithmically spaced energy grid with width
0.025 dex from 0.3 to 6.5 keV.  

Background at the position of each source is evaluated by normalizing
an average background at the source off-axis to the source spectrum
accumulated between 7 and 11 keV, where the contribution of faint
sources is negligible, due to the sharp decrease of the mirror
effective area. We do not use a 'local' background for two main
reasons: 1) minimize the systematic error due to sharp background
variations and 2) search for source detections in relatively narrow
energy bands, where the 'local' background counts may be small and the
associated statistical uncertainty comparatively high.

\subsection{Background evaluation}

\begin{figure}[h]
\centering
\includegraphics[width=6.cm,angle=-90]{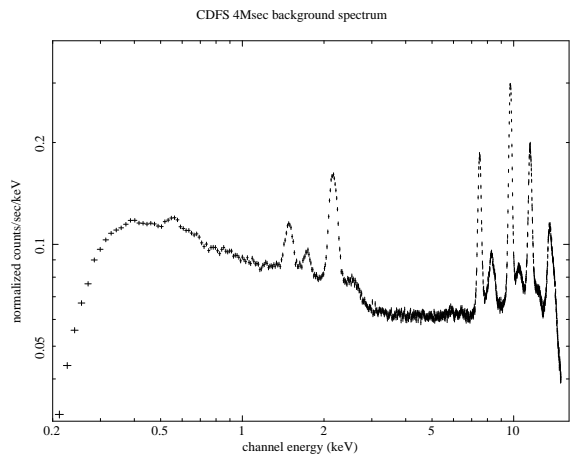}
\caption{The spectrum of the background in the Chandra 4 Msec exposure
  of the CDFS.  The spectrum has been extracted from a circle of 8
  arcmin radius, after the exclusion of 5 arcsec radius region around
  all detected sources (10-20 arcsec for bright sources and extended
  sources).}
\label{bgd}%
\end{figure}

The study of faint X-ray sources requires the best possible
characterization of the background. The typical level of the CDFS
background in the 0.5-7 keV band is 2-3 counts/arcsec$^2$ or
$\sim$25-40 counts in an area of 2 arcsec radius. The spectrum of the
Chandra background is complex, with several broad and narrow
components, both internal and of cosmic origin.  To characterize the
Chandra background we extracted its spectrum from the following regions:
1) circular region with radius 8 arcmin; 2) circular region with
radius 3 arcmin; 3) annulus with internal radius 3 arcmin and external
radius 6 arcmin; 4) annulus with internal radius 6 arcmin and external
radius 8 arcmin.  Circular region of 5 arcsec radius around all X-ray
detected sources have been excluded from the analysis
in all cases. Fig. \ref{bgd} shows the spectrum extracted from the broad 8
arcmin radius region. The continuum above a few keV is dominated by particle
induced background (PIB), which has a power law shape with {\it
  positive} spectral index above 4-5 keV.  Over-imposed to this
continuum we see at least 8 strong emission lines.  The best fit
energies of the lines are and their identification is given in Table A.1.
The rising background continuum and the strong emission lines present
above $\sim 7$ keV, coupled with the strongly decreasing mirror
effective area above this energy, suggest to limit the analysis of
faint sources to the band below 7 keV.

\begin{table}
\caption{\bf Chandra background lines}
\begin{tabular}{lcc}
\hline
Energy    & Width  & identification \\
keV       & keV    & \\
\hline
0.585    & 0.05  & Blend of Fe-L and O lines\\
1.49     & 0.015 & 1.4867 Al K$\alpha$(+1.5575 Al K$\beta$) \\
1.78     & 0.010 & 1.7400 Si K$\alpha$ (+1.8359 Si K$\beta$) \\
2.16     & 0.045 & Au M$\alpha \beta$ \\
7.48     & 0.05  & 7.47 Ni K$\alpha$ \\
8.26     & 0.16  & 8.28 Ni K$\beta$ \\
9.71     & 0.06  & 9.71 Au L$\alpha$  \\
11.5     & 0.1   & 11.44 Au L$\beta$  \\
13.6-14  & 0.2   & 13.38 Au L$\gamma$  \\
\hline
\end{tabular}
\end{table}

The low energy background spectrum is peaked at $\sim 0.5$ keV and
decreases rather sharply between $\sim 0.6$ and $\sim 1$ keV. This
continuum and the 0.585 line suggest that this part of the background
spectrum is dominated by the thermal emission from the local
``superbubble''. Below 0.3 the response of the instrument decreases
sharply and therefore we limit the analysis at energies above 0.3 keV.

We have then fitted the background spectrum between 0.3 and 7 keV with
a model including: 1) three gaussian lines to account for the features
at 1.48, 1.74 and 2.16 keV; 2) one broad gaussian line to reproduce
the broad bump between 1 and 3 keV; 3) two power laws modified at low
energy by photoelectric absorption to reproduce the broad band hard
continuum. All these components are not convoluted with the mirror
effective area. 4) a thermal component with abundances fixed to Solar
value, convoluted with the mirror effective area. This model provides
a good fit to the background spectrum, without systematic
residuals. We have then fitted the same model to the background
spectra accumulated in three different regions, adjusting only the
global normalizations of the model spectra accounting for the internal
and cosmic background components.  The fits are remarkably good above
1 keV. Below this energy we see a deficit of counts in the 3 arcmin
radius spectrum and in the 3-6 arcmin radius annulus spectrum, while
there is an excess in the 6-8 arcmin annulus spectrum. The spectra
become increasingly soft at large off-axis angles. The
relative intensity of the thermal component is higher in the on-axis
spectrum than in the off-axis spectrum, as expected if this component
is truly cosmic, because of the larger mirror vignetting at high
off-axis angles.  The increase of the strength of the low energy
cut-off at smaller off-axis is then probably due to a thicker
deposition of contaminants at the centre of the ACIS-I 4 chip region.

The behavior observed for the emission lines is quite
interesting. While the intensity of the 1.49 keV and 2.16 keV
lines is constant with the off-axis angle, the intensity of the 1.78
keV Si line is larger at large off-axis angles.

The variations of the background spectra accumulated in thinner annuli
becomes comparable with the statistical uncertainty.  Guided by this
analysis we decided to adopt the average background spectra
accumulated in the three $r<3'$, $3'<r<6'$ and $6'<r<8'$ in the
following analysis.

To evaluate the background counts in each source extraction region we
need to normalize the average background at three off-axis angles to
that at each source position. To this purpose we first extract spectra
from circular regions of 10 arcsec radii centered at each galaxy
positions, and then normalize the average background to that in these
spectra using the 7-11 keV faint-source-free band. The statistical
uncertainty on the background evaluation is therefore mostly given by
the 7-11 keV total number of counts at the source position. The choice
of the size of the extraction region at each galaxy position must then
satisfy two competing requirements. On one hand the region must be
large enough to provide enough counts to keep statistical fluctuation
small. On the other hand, a small region has the advantage of avoiding
systematic errors due to sharp variations of the exposure time (and so
of the background) typical of mosaics like the ACIS-I CDFS exposures,
cosmetic defects, crowded regions with several sources within few
arcsec.  The average CDFS 7-11 keV background in a 10 arcsec radius
region is between 600 and 700 counts,
$\sim3$ times higher than the 0.5-2 keV average
background, 4 times the 1-2 keV background and 2 times the 1-4 keV
background. Therefore, a circular region of 10 arcsec radius is a good
compromise, because it collects a number of 7-11 keV counts similar or
higher than the 0.5-2, 1-2 or 1-4 keV counts in 10-20 arcsec annulus
(a typical region used to extract a 'local' background in crowded
areas like the CDFS), in a three time smaller area.

\subsection{Completeness, reliability and sky coverage}

To extract statistically quantitative information from a given sample
of ``detected'' sources we need to first assess its completeness and
reliability (number of spurious detections).  To this purpose we
performed a series of extensive detection runs on simulated data.

First, to assess the reliability of our survey we performed
simulations using background spectra only. We simulated about $10^5$
spectra at the positions of the sources detected in the CDFS to use
exactly the same exposure time, vignetting and PSF, but including only
the average background at each off-axis position. We run our {\it
  ePhot} on these simulations and studied the number of spurious
detections as a function of various parameters: 1) the threshold on
background subtracted counts in a wide (0.3-4 keV) band; 2) the
minimum and maximum source extraction regions; 3) the energy grid used
to search for detections; 4) the band-width $E_{max}/E_{min}$; 5) the
Poisson probability threshold for a source detection. We choose the
combination of parameters that keep the number of spurious detections
$\ls1$ every 5000 spectra, i.e. the number of candidates in our search
(see Sect. 2): 0.3-4 keV background subtracted counts $\ge 12$,
$E_{max}/E_{min}>2$, Poisson probability $<2\times10^{4}$.

\begin{figure}[t!]
\centering
\includegraphics[width=8cm]{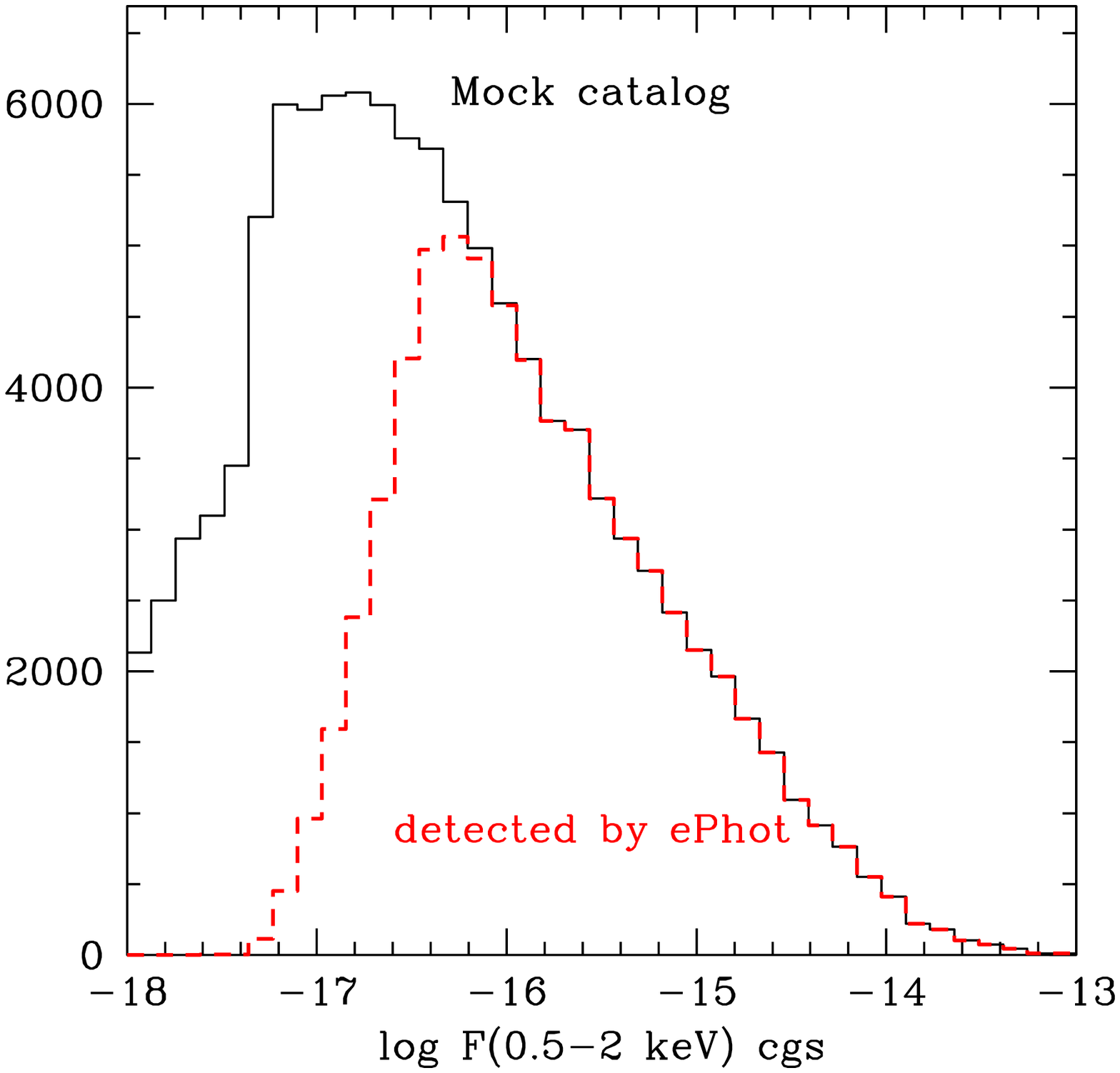}
\includegraphics[width=8cm]{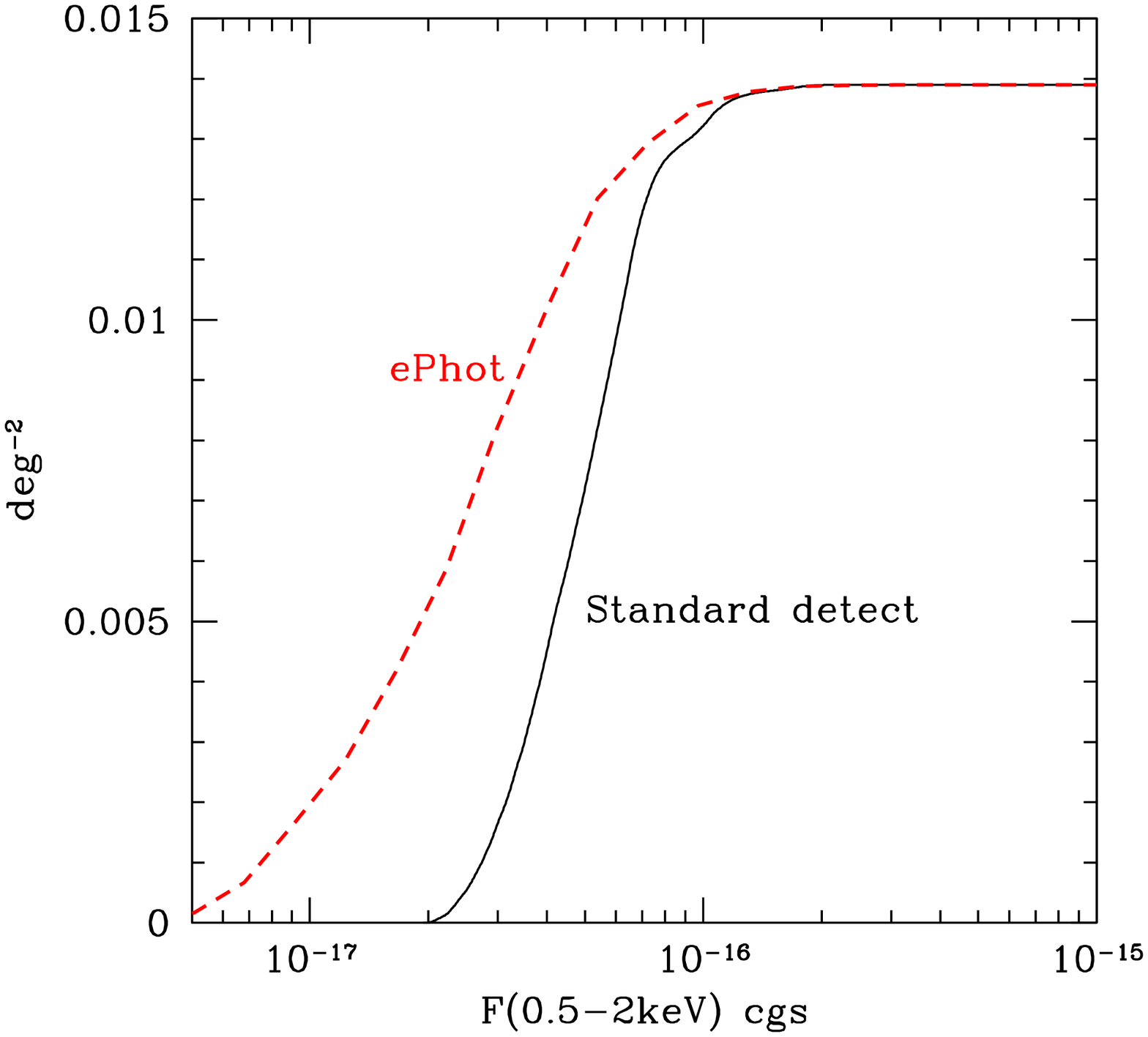}
\caption{[Top panel:] the 0.5-2 keV flux histogram of the simulated
  sources (black solid histogram) and of the simulated sources
  detected by {\it e-phot}. [Bottom panel:] the Chandra sky coverage
  in the ERS area obtained using a standard detection algorithm (black
  solid curve) and using {\it e-phot} (red dashed curve) at the same
  probability threshold.}
\label{simul}%
\end{figure}

To study the completeness of our survey we performed detection runs on
simulated source plus background spectra. We used the Mock X-ray
catalogs generated by R. Gilli, including about $10^5$ AGN with a
distribution of luminosities, redshifts and absorbing column densities
in agreement with the \cite{gilli:2007} CXB model.  In particular, the
Mock catalogs include sources with unabsorbed flux down to $10^{-18}$
cgs, where the AGN density is $\sim20000$ deg$^2$. Source spectra are
distributed over six templates, with column densities N$_H$ up to
$3\times10^{24}$ cm$^{-2}$. A pure reflection spectrum is also
considered to account for heavily Compton thick sources. The
distributions of spectra is a strong function of both luminosity and
redshift (see \cite{gilli:2007} for details).  Using these Mock
catalogs we simulated a 10 $deg^2$ Chandra survey with the same
characteristics of the CDFS survey. We simulated sources at the
position of the sources detected in the CDFS. We added to each
simulated spectrum a background proportional to the average background
at each off-axis angle. Finally, we run our on each source +
background simulated spectrum our source detection algorithm {\it
  ePhot}. These simulations are used to study the completeness of our
survey and compute the corresponding sky coverage.  Fig. \ref{simul},
top panel, compares the 0.5-2 keV observed flux histogram of the
simulated sources with that of the sources detected by {\it ePhot}
with the thresholds defined above. The ratio of the two histograms
gives the completeness of our survey, or the sky coverage, when
normalized to the CDFS searched area. Fig. \ref{simul}, bottom panel,
compares the {\it ePhot } sky coverage for the ERS area of the CDFS to
that of a traditional detection algorithm at the same probability
threshold (same fraction of spurious detection).

\subsection{Photometry}

{\em e-phot} was also run on the galaxy samples with fixed energy
bands (0.5-2 keV, thus optimizing only for the size of the source
extraction region).  X-ray fluxes in the band 0.5-2 keV are estimated
from {\em e-phot} count rates in the 0.5-2 keV band if the signal to
noise ratio in this band is higher than 2.5 or from the count rates in
the band that optimize the detection otherwise.  A simple power law
spectrum with $\alpha_E=0.4$ has been assumed for the count rate flux
conversion.  To test our photometry we compared the 0.5-2 keV flux
obtained from the event files of the first 2 Msec observations to
those in Luo08, and the 0.5-2 keV flux obtained form the full 4 Msec
observations to those in Xue11.  The agreement is remarkable,
the median of the two samples agree within 3\% and the
semi-interquartile range of their ratio is 0.06.  Fluxes of faint
detections can be statistically over-estimated because of the
so-called Eddington bias (\cite{hogg:1998,wang:2004}). We estimated
the bias according to \cite{wang:2004} and found that it is at most
$\sim20\%$ for our faintest detection and it is negligible for most of
the Chandra-GOODS-ERS and Chandra-GOODS-MUSIC sources. Luminosities in
Table 3 are calculated from fluxes corrected for the Eddington bias.

{\bf Acknowledgements}

This work was supported by ASI/INAF contracts I/024/05/0 and
I/009/10/0. FS acknowledges support from the Alexander von Humboldt
Foundation.  We thank R. Gilli for providing us with his mock catalogs
of X-ray sources based on the GCH2007 model.  FF thanks G.C. Perola,
A. Comastri, M. Brusa, F. La Franca, G. Melini, R. Maiolino,
E. Giallongo and H. Netzer for useful discussions.  This work is based
on observations made with NASA X-ray observatory Chandra. We thank the
Chandra Director's office for allocating the time for these
observations. X-ray data were obtained from the archive of the Chandra
X-ray Observatory Center, which is operated by the Smithsonian
Astrophysical Observatory.  This work is also based data obtained with
the NASA/ESA Hubble Space Telescope, obtained from the data archive at
the Space Telescope Institute. STScI is operated by the association of
Universities for Research in Astronomy, Inc. under the NASA contract
NAS 5-26555.  Observations were also carried out using the Very Large
Telescope at the ESO Paranal Observatory under Programme IDs
LP181.A-0717, ID 170.A-0788, and the ESO Science Archive under
Programme IDs 67.A-0249, 71.A-0584, 69.A-0539.

\end{document}